\documentstyle[a4,12pt]{article}
\renewcommand{\baselinestretch}{1.5}
\begin{document}
\newcommand{\C}{C^{z}}
\newcommand{\Hb}{H^{z}_{\bar{\theta}}}
\newcommand{\D}{D_{\theta}}
\newcommand{\Db}{D_{\bar{\theta}}}
\newcommand{\zet}{\zeta_{\theta}}
\newcommand{\der}{\partial_{z}}
\newcommand{\R}{R_{z\theta}}
\newcommand{\covder}{\hat{\nabla}_{\zeta}}
\newcommand{\Del}{\Delta_{\zeta}}
\newcommand{\Delk}{\Delta_{\chi}}
\newcommand{\Delko}{\Delta_{\chi_{0}}}
\newcommand{\F}{\Phi_{\zeta}}
\newcommand{\Fk}{\Phi_{\chi}}
\newcommand{\Fko}{\Phi_{\chi_{0}}}
\newcommand{\zetp}{\zeta^{\prime}}
\newcommand{\demi}{\frac{1}{2}}
\newcommand{\dl}{d\lambda}
\newcommand{\dlb}{d{\bar{\lambda}}}
\newcommand{\M}{d^{2}\lambda}
\newcommand{\sig}{\hat{\Sigma}}
\newcommand{\dsig}{\partial\hat{{\cal D}}(\eta_{0})}
\newcommand{\fact}{\frac{1}{2i\pi}}
\newcommand{\facth}{\frac{1}{8i\pi}}
\newcommand{\kio}{\chi_{0}}
\newcommand{\Rko}{R_{\chi_{0}}}
\newcommand{\Tb}{\bar{T}_{\bar{\lambda}}}
\newcommand{\Pz}{P_{0k}}
\newcommand{\zt}{\tilde{z}}
\newcommand{\tetil}{\tilde{\theta}}
\newcommand{\Zh}{\hat{Z}}
\newcommand{\Teth}{\hat{\Theta}}
\newcommand{\zett}{\zeta_{T}}
\newcommand{\Pb}{\bar{P}}
\newcommand{\Dec}{\bigtriangledown}
\renewcommand{\baselinestretch}{1.2}
\def\scf{superconformal }
\def\scfty{superconformality }
\def\sd{superdiffeomorphism }
\def\sc{supersymmetric }
\def\sdf{superdifferential }
\def\sdfs{superdifferentials }
\def\sz{supersymmetrization }

\pagestyle{plain}
\date{}
\title{\bf N=1, 2d-induced Polyakov supergravity on a super Riemann
surface }
\author{{\bf H. Kachkachi $^{\dag}$}\\
{\em Laboratoire de Physique Th\'{e}orique,$^{\ddag}$}
{\em Universit\'{e} de Bordeaux I, }\\
{\em 19 rue du Solarium, F-33175 Gradignan Cedex, France.}}
\begin{titlepage}
\maketitle
\thispagestyle{empty}
{\centerline{\bf Ph.D. Thesis}}
\begin{abstract}
An effective action describing N=1, 2d-induced supergravity in the chiral
gauge is obtained on the supertorus and on a compact SRS without boundary of
arbitrary genus. This action integrates the superconformal Ward identity for
the superdiffeomorphism anomaly; it yields the last step in the  extension
to more complicated topologies of the action initially obtained by A.M.
Polyakov on the ordinary complex plane. Constructing the Weierstrass $\zeta$
function as the Cauchy kernel on the supertorus allows for solving the super
Beltrami equations and thereby for computing the stress-energy tensor and
2- and 3-point Green functions. Using this result one checks the Polyakov
conjecture which states that the Polyakov action resums the renormalized
perturbative series. The stress-energy tensor is also constructed on a SRS.
Obtaining these results was made possible by analysing new geometrical
notions such as single-valued superdifferentials via the super Riemann-Roch
theorem, multi-valued superdifferentials, a non-Berezinian method for
integrating Grassmann variables, super Stokes theorem, Cauchy kernels and
covariant derivatives.
\end{abstract}
Email address: {\bf Kachkachi@FRCPN11.IN2P3.FR}\\
$\dag$ $\overline{\mbox{Unit\'{e} Associ\'{e}e au CNRS}, U.A.764}$\\
\end{titlepage}
\section{Introduction}
For decades high energy physicists have made a tremendous effort in
an attempt to incorporate the gravitational force into the general scheme of
renormalizable gauge theories, the so-called Grand Unified Theory which
includes the electromagnetic, weak and strong interactions.
The fundamental obstacle to this aim is the fact that gravity has a coupling
constant that, owing to the equivalence principle, is dimensionful. This means
that gravity escapes the usual renormalization program of gauge field
theories, as one cannot add diagrams with different powers of the
coupling constant.
Then, what is needed is a field theory that can describe both gravity and
the other interactions and which is based upon the principles of special and
general
relativity and of quantum mechanics, and that, though unrenormalizable, still
has a predictive power. At present, it is believed that the most promising
hope to realize this program is held by superstring theory. This is a theory
of one-dimensional extended objects, which, when moving in spacetime, sweep
out a two-dimensional surface, the worldsheet. Superstring theory is based on
a new symmetry principle, namely a symmetry between bosonic and fermionic
fields called supersymmetry. It is a remarkable fact that supersymmetry can be
implemented in field theory if spacetime is curved and hence if gravity is
present. In fact, it turns out that superstring theory not only includes
Einstei
general relativity and Yang-Mills theory, but it also includes supergravity
and the Grand Unified Theory.\\
In supergravity, the symmetry between bosons and fermions implies the
existence of one fermionic companion to the usual bosonic gravitational field,
the graviton, and other fermionic companions to the other bosonic fields. The
fermionic companion to ( or the super-partner of ) the graviton is a spin
$\frac{3}{2}$ field, called the gravitino. It has been shown that the only
consistent field theory of interacting spin $\frac{3}{2}$ fields is
supergravity. Moreover, because of the short-range nature of the exchange of
fermionic fields, supergravity differs radically from ordinary general
relativity at short distances. In particular, due to the symmetry between
bosons and fermions, infinities ( in the first and second order ) in the
S-matrix of supergravity cancel out. Thus supergravity could be regarded as
having such a predictive power as ordinary renormalizable models, and thence
provides a candidate for the sought-for theory.\\
There are several models of supergravity characterized by the number $N$ of
supersymmetries. However, it was shown that these N-extended supergravities
are only viable for $0\leq N \leq 8$. In this work, however, we will only be
concerned with the simple ($N=1$) supergravity which contains only one
graviton-gravitino multiplet together with other pairs of bosons and their
fermionic partners of lower spins.\\
A consistent framework for studying $N=1$ supergravity is provided by the
covariant Ramond-Neveu-Schwarz model of the superstring. In this theory
space-time Lorentz invariance is manifest, while space-time supersymmetry is
not. \\
In this model, the evolution of a superstring
in space-time can be parametrized by a pair consisting of a bosonic scalar
field $X^{\mu}$ ( position of the bosonic string ) and a spin $\frac{1}{2}$
fermionic field $\psi^{\mu}$. These define smooth embeddings of the worldsheet
$\hat\Sigma$ swept out by the superstring into space-time. These fields can be
treated as a boson and its fermionic partner
in a multiplet called a superfield, if we introduce an anticommuting complex
variable $\theta$ in addition to the usual complex variable $z$, so that a
point on the worldsheet $\hat\Sigma$ is parametrized by the pair of
coordinates $(z,\theta)$. Thus the position vector of the superstring in
space-time is a superfield that can be written as $\Phi^{\mu}(z,\theta)=
X^{\mu}(z)+\theta\psi^{\mu}(z)$. Then, superstring theory can be formulated
as the superfield $\Phi^{\mu}(z,\theta)$ coupled to 2-dimensional $N=1$
supergravity, i.e. to the superfield corresponding to the graviton-gravitino
multiplet. The latter is called the supervielbein ( or the superzweibein in
2-dimensions ) and defines the geometry of
the corresponding supergravity theory, just as the vielbein defines the
geometry of space-time in general relativity.\\
It is known \cite{Howe} that any supergravity geometry in two dimensions is
locally
equivalent to a flat geometry by  superconformal transformations of the
coordinates $(z,\theta)$. This means that there exist local coordinates in
which the superzweibein becomes flat. These local complex coordinates
$z=\sigma_{1}+i\sigma_{2}, \theta=\theta_{1}+i\theta_{2}$,  together with
superconformal transformations ( transition functions ) define a superconformal
manifold of complex dimensions $(1\vert 1)$, called a super Riemann surface
(SRS).
Then, when we consider interactions at a given loop order $g$, the worldsheet
of the superstring is a compact SRS of genus $g$, this is what we denoted
above by $\hat\Sigma$. The corresponding action has a large
gauge symmetry: It is invariant under reparametrizations or diffeomorphisms
of $\hat\Sigma$ and their supersymmetric partner, the local supersymmetry
whose corresponding gauge field is the gravitino, it is also invariant under
superconformal rescalings or super Weyl transformations as well as under the
local Lorentz transformations of the superzweibein.

At the quantum level, superstring theory exists in two entirely different
versions: in the form of canonical quantization it appears as the
representation theory of the algebras of Heisenberg, Virasoro, and Kac-Moody
and their supersymmetric extensions. In the second version, quantization
is performed with the help of the Polyakov path formalism in which one sums
over all random surfaces \cite{Pol1}. The latter leads to
the analytic theory of algebraic supercurves and their moduli spaces. In the
operator formalism one finds it difficult to construct a consistent picture of
interacting strings because of the infinitesimal (though mathematically
rigorous
aspect of this formalism. The Polyakov formalism is, on the contrary, geometric
and can thus treat global objects, though it suffers the shortcoming of lacking
a rigorous mathematical formulation.\\
In the Polyakov formalism, that will be adopted in this work, quantization
involves
functional integration over the superfield $\Phi^{\mu}$ and the superzweibein.
Since the integral over $\Phi^{\mu}$ is gaussian, only the integration over
the superzweibein is non--trivial. In order to avoid the overcounting of
equivalent configurations following from the fact that supergravity geometries
related by superdiffeomorphisms and superconformal transformations define the
same SRS, we must pick a gauge-fixing slice transverse to the orbits of the
super Weyl and superdiffeomorphism groups in the space of superconformal
equivalence classes of metrics, i.e. a slice that contains
one representative of each equivalence class, and then integrate over it.
The space of these equivalence classes is called the supermoduli space
of super Riemann surfaces. This is a supermanifold of complex dimensions
$(3g-3\vert 2g-2)$ when
$g\geq 2$, with the moduli space as its underlying manifold, see sect.1.2 for
details. Thus in the Polyakov path formalism, the functional integral reduces
to a finite integral over the supermoduli space of SRS's.
In this setting, scattering amplitudes of the string are integrals of various
functional determinants (Laplacian determinants) over supermoduli space. The
Laplacian determinants turn out to be products of first order operators which
depend holomorphically on moduli.\\
A powerful tool for studying the holomorphic structure of the string
scattering amplitudes in such a representation, their finiteness, and even to
compute them in terms of theta functions via the Selberg trace formula
\cite{BMFS,Grosch}, is provided by its underlying fundamental symmetry,
namely the superconformal symmetry of the worldsheet, and its relation to
algebraic geometry. Indeed, conformal invariance by itself is sufficient to
determine almost completely the structure of $N-$point Green functions
by only stressing the group theoretic behaviour of background fields under
conformal transformations.

Let us go back to the problem of functional integration and note that the
non-trivial integration over the superzweibein leads to two different settings
depending on the gauge slice of superzweibeins we choose. In the
superconformal gauge obtained after transforming the superzweibein by
superdiffeomorphisms and super Weyl rescalings into a flat reference
superzweibein, the functional
integration analysis leads to the so-called super Liouville theory
\cite{DisKa}. This is
a two-dimensional field theory of the scalar field of Weyl
rescalings, called the Liouville mode. This represents the degree of freedom
of the 2-dimensional supergravity. The corresponding action yields a
measure of the violation of the conformal symmetry at the quantum level.
However, this symmetry is restored in some dimensions of space-time, 26 for
bosonic and 10 for fermionic strings, the so-called critical dimensions. This
follows from the fact that in these particular dimensions the dependence on
the Liouville mode disappears from the path integral as this contains an
explicit factor that vanishes at these dimensions \cite{Pol1}. Therefore, to
quantize the string in the conformal gauge away from the critical dimensions,
it is necessary to solve the Liouville action exactly.\\
One can, instead of using the conformal gauge, choose the light-cone or
chiral gauge, which has a single non-vanishing metric mode, the super
Beltrami differential that represents the graviton-gravitino multiplet, and
recast the theory in a local form by introducing the Wess-Zumino field defined
by the super Beltrami equation. In mathematical terms, this field is the
projective coordinate that represents the isothermal (or projective) structure
parametrized by the super Beltrami differential. The resulting effective
action is a Polyakov action that describes the Wess-Zumino field, and
is invariant under reparametrizations, while its variation under Weyl
rescalings produces the conformal anomaly.\\
This gauge-fixing scheme can be better understood if one makes use of
conformal field theory on super Riemann surfaces. A precious
result that stems from the combination of algebraic
geometry of SRS and superconformal field theory thereon, is that in critical
dimentions, the quantum functional measure is basically the square
modulus of a holomorphic function on moduli \cite{BMFS,BM,BK,Grosch}.
Unfortunately, one can simultaneously maintain, upon quantization, holomorphy
and reparametrization invariance only in critical dimensions where
all local anomalies cancel out.
Nevertheless, off critical dimensions the Weyl anomaly can be changed to a
diffeomorphism anomaly by extracting from the effective action a suitable
local counterterm \cite{KLT, AGN}. This enables us to exploit the salient
feature of the diffeomorphism anomaly namely, the holomorphic factorization,
which is thus restored.\\
The holomorphic factorization consists in separating those correlation
functions which depend on the projective coordinates from those which depend
on their complex conjugates.
But since these coordinates, being solutions of the Beltrami equations, are
holomorphic functionals of the Beltrami
differential, the holomorphic factorization can be recast into separating
functionals of the Betrami differential and of its complex conjugate.\\
Since the holomorphic factorization is an important property of the action
we are going to construct, it is necessary to distinguish the bosonic from
the supersymmetric case. Although our work is only concerned with
$N=1$ induced supergravity, the discussion of the bosonic case will serve
as an introduction to the corresponding development in next chapters, as this
constitutes the starting point for our supersymmetric generalizations.

In the bosonic case, if we denote the Beltrami differential by $\mu$,
the holomorphic factorization consists in splitting the Weyl invariant
effective action into a chiral (holomorphic) and anti-chiral
( anti-holomorphic ) terms
\begin{equation}\label{intra}
\Gamma_{P}[\mu,\bar\mu;R_{0},\bar R_{0}] = \Gamma_{P}[\mu;R_{0}] +
\overline{\Gamma_{P}[\mu;R_{0}]}
\end{equation}
where $R_{0}$ is a background holomorphic projective connection in
the conformal reference structure $\{(z,\bar z)\}$, i.e.
$\bar\partial R_{0}\equiv\partial R_{0}/\partial\bar{z}=0$; it is introduced
so as to make the diffeomorphism anomaly well-defined.
The chiral functional on the right-hand side of (\ref{intra}) is a functional
which depends holomorphically on the background conformal geometry
parametrized by the pair $(\mu, R_{0})$. This functional, called the induced
Polyakov action, serves as a ``classical''
action for 2-D quantum gravity in the so-called light-cone gauge, i.e.
$ds^{2}=(dz+\mu d\bar{z})d\bar{z}$. This
Polyakov action $\Gamma_{P}[\mu; R_{0}]$ satisfies the chiral conformal Ward
identity,
\begin{equation}\label{intrb}
s\Gamma_{P}[\mu; R_{0}]= k {\cal A }(c;\mu;R_{0})
\end{equation}
with ${\cal A }(c;\mu;R_{0})$ the well-defined ( or globally defined, which
means that it remains invariant under a holomorphic change of coordinates
on a Riemann surface ) integrated diffeomorphism
anomaly which is a solution of the Wess-Zumino consistency condition,
$s{\cal A }(c;\mu;R_{0})=0$; its expression is given below by eq.(\ref{chiig}).
Here $s$ is the BRST operator associated with the
diffeomorphism group, and is obtained by replacing the diffeomorphism
infinitesimal parameters $\xi$ and $\bar\xi$ by the Faddev-Popov ghost $\gamma$
and its complex conjugate $\bar\gamma$ respectively; the diffeomorphism ghost
field $c$ is defined by $c=\gamma+\mu\bar\gamma$ \cite{Bec}. $k$ is the
central charge of the model and measures the strength of the diffeomorphism
anomaly; it is
proportional to the central charge of the Virasoro algebra generated by the
energy-momentum tensor of the original matter system. The anomaly
${\cal A }$ represents the center-extension cocycle of the Virasoro algebra.\\
The fact that this action depends only on the background conformal
geometry suggests that the study of 2-dimensional conformal field theories on
Riemann surfaces should rely on conformal geometry, and thus a
starting point for this study is to solve the conformal Ward identity
(\ref{intrb}). Accordingly, a unique solution on the complex plane was
found by Polyakov in \cite{Pol2}, the solution on the torus was constructed by
Lazzarini and Stora in \cite{Laz}, and quite recently Zucchini has found the
generalization of these solutions to a Riemann surface of higher genus
\cite{Zu1}. In this case the solution is non-unique, since it is only defined
up to addition of an arbitrary local
holomorphic function due to the presence of zero modes. We will consider these
solutions in more detail later on.

Now we come to discussing the supercase and defining the framework for our
purpose.
The Polyakov action for the effective induced
supergravity in two dimensions can be defined in analogy with the bosonic
case. In order to write the holomorphic factorization of the
superdifferomorphism anomaly, let us first define the corresponding
geometrical setting. Here we consider a compact SRS $\hat\Sigma$ (without
boundary) of genus $g$, with a reference conformal structure $\{(z,\theta)\}$
together with an isothermal structure $\{(\hat{Z},\hat\Theta)\}$, this is
obtained from the reference one by a quasisuperconformal transformation, i.e.
a transformation that changes a circle into an ellipse. This
transformation is parametrized in general by three super Beltrami
differentials of which only two are linearly independent. There is a
formalism in which one of the independent differentials is set to zero as it
contains only non-physical degrees of freedom, thus ending up with only one
super Beltrami differential. Setting to zero this differential implies the
existence of a superconformal structure on the SRS which is necessary, on the
other hand, for
defining the Cauchy-Riemann operator. This formalism is used in the papers
{\bf II} and {\bf IV}
\footnote{Here and later on we use {\bf I, II, III, IV} to refer to our
respective papers \cite{HKMK1}, \cite{AK1},  \cite{HKMK2}, \cite{AK2}.}.
However, it is more natural from a geometrical point of view to work in
another formalism that also reduces the number of super Beltrami differentials
to one by eliminating the $\bar\theta-$dependence in the coordinates
$(\hat{Z},\hat\Theta)$ \cite{CraBin1,Tak} in addition to the
superconformal-structure condition of the previous formalism. The
superconformal
thus defined is parametrized by a single super Beltrami differential
$\hat\mu$, see sect.1.3 for details. In this particular gauge of super
Beltrami differentials, that we call the Beltrami gauge, the absence of
$\bar\theta-$dependence in the coordinates $(\hat{Z},\hat\Theta)$ implies
that the action we are going to construct actually describes the
$(1,0)-$ supergravity of the graviton and gravitino fields contained in
$\hat\mu$. More importantly, this gauge allows for decoupling the super
Beltrami
equations satisfied by $\hat{Z}$ and $\hat\Theta$ which are thus more easily
solved by
using the techniques of the Cauchy kernel. The solutions thus obtained enable
us to write the action as a functional of the super Beltrami differential
$\hat\mu$ and then to compute the Green functions and the energy-momentum
tensor whose external source is $\hat\mu$.
Later on, we will see that the coordinate $\hat\Theta$ is the
Wess-Zumino field introduced above in the supersymmetric case.\\
In this parametrization, the super Weyl invariant effective action splits
into two terms, i.e.
\begin{equation}\label{intrc}
\Gamma\!_P [\hat\mu,\bar{\hat\mu}; \hat R_0,\bar{\hat R}_0]=
\Gamma\!_P [\hat\mu;\hat R_0]+\overline{\Gamma\!_P [\hat\mu;\hat R_0]}
\end{equation}
where $\hat R_0$ is a holomorphic background superprojective connection in
the superconformal structure $\{(z,\theta)\}$, i.e.
$\bar{D_{\theta}}\hat R_{0}=
(\partial_{\bar\theta}+\bar{\theta}\partial_{\bar z})\hat R_{0}=0$.
For the same reason as in the bosonic case,
$\hat R_{0}$ is introduced to insure a good glueing of the anomaly on
$\hat\Sigma$.\\
The chiral part on the right-hand side of eq.(\ref{intrc}) is the 2-d induced
Polyakov action that describes quantum supergravity in the light-cone
gauge, i.e. $ds^{2}=(dz+\hat{\mu}d\bar{z}+\theta d\theta)d\bar{z}$
. It depends on the background conformal geometry parametrized by the
pair $(\hat\mu,\hat R_0)$.
This action satisfies the superconformal Ward identity \cite{Grixu,DelGie}
\begin{equation}\label{intrd}
(\bar\partial -\hat\mu\partial -{3\over 2}\partial\hat\mu -{1\over 2}D\hat\mu
D){\delta\Gamma\! \over \delta\hat\mu} =\kappa\partial^2D\hat\mu\ .
\end{equation}
which is also the non-holomorphy equation of the stress-energy tensor.
Using the BRST operator $s$ and the well-defined anomaly, we equivalently
have

\begin{eqnarray}\label{intre}
s\Gamma\! [\hat\mu,\hat
R_0]&=&\kappa\int_{\hat\Sigma}d\tau[C\partial^2D\hat\mu-
\hat\mu\partial^2 DC) +3\hat R_0(C\partial\hat\mu-\hat\mu \partial C)+
D\hat R_0(CD\hat\mu-\hat\mu DC)] \nonumber\\
&\equiv& \kappa {\cal A}(C,\hat\mu,\hat{R}_{0}),
\end{eqnarray}

with $d\tau\equiv\frac{dz\wedge d\bar{z}}{2i\pi}d\theta$.
Here $C$ is the superdiffeomorphism ghost
field; $\kappa$ is the central charge of the model and measures the strength
of the superdiffeomorphism anomaly; as in the bosonic case this is the only
remnant of the matter system after functional integration.
${\cal A}(C,\hat\mu,\hat{R}_{0})$ is the well-defined integrated anomaly on
the super Riemann surface (SRS) $\hat\Sigma$. \\
Solving Eqs.(\ref{intrd}) or (\ref{intre}) on a super Riemann
surface of genus $g\geq 0$ is, as in the bosonic case, the
starting point for studying 2-dimensional superconformal models thereon.
A solution to this superconformal Ward identity was found by Grundberg and
Nakayama in \cite{GrunNak} on the super complex plane.
Then generalizations of this solution to the supertorus and to a SRS of higher
genus, which is the subject of this work, have been performed by the author
in collaboration with J.-P. Ader in the papers {\bf II} and {\bf IV}.
Now I come to present the program of this work in more detail.\\

In chapter I, I first recall in a self-contained way the basic features of
the theory of SRS's which are of direct use in our work, and also present
some new relevant points I have developed in the papers {\bf I} and {\bf III}
in collaboration with my brother M. Kachkachi on which paper {\bf IV} is
based. Accordingly, I will present a
SRS $\hat\Sigma$ as having two superconformal (reference and isothermal)
structures. The isothermal structure will be parametrized by the
supercoordinates $(\hat Z,\hat\Theta)$ that we shall regard as being obtained
from the reference structure coordinates $(z,\theta)$ by performing a
quasisuperconformal transformation parametrized by the single super Beltrami
differential $\hat\mu$ in the Beltrami gauge. In this gauge
the super Beltrami equations are easily decoupled, and then to solve them
we construct in ({\bf III}) the
quasielliptic Weierstrass $\zeta$-function as the $\bar\partial$-Cauchy
kernel on the supertorus. We also suggest an analogous object
on a SRS; however in this case much is still needed about, for instance, the
boundary behavior of this prime form in order to solve the super Beltrami
differentials (SBE) thereon.\\
Next we study in detail the monodromy properties of multi-valued
superdifferentials as distinguished from single-valued ones \cite{HK}.
The monodromy of a multi-valued ( or polydromic ) field is defined as the
phase factor that this field picks up when it is carried along a close curve
on a Riemann surface that encircles a singular point of the field.
Multivalued differentials are no-where vanishing and transform up to a
character under the group of superdiffeomorphisms of the SRS \cite{HK}.
On the other hand, the Riemann-Roch theorem states that an even
holomorphic single-valued $\frac{1}{2}-$superdifferential possesses globally
( and counting multiplicity ) $(g-1)$ zeros on a compact super Riemann surface
of genus $g$. In {\bf I} we use this theorem to give a local expression for a
holomorphic (single-valued) $\frac{1}{2}-$superdifferential in a
superconformal structure parametrized by special isothermal coordinates on a
SRS; this expression is very crucial for carrying explicit residue calculus.
The holomorphy of these coordinates with respect to super Beltrami
differentials is proved. This property inssures the holomorphic factorisation
of the effective action as in eqs.(\ref{intra}), (\ref{intrc}) above.\\
Out of these differentials we first construct superaffine and superprojective
connections, then the corresponding supercovariant derivatives. These will
prove to be of practical necessity to build globally defined Lagrangian
densities in the Polyakov formalism. Then with all this material at hand,
we will proceed to the construction of the Polyakov action. \\
We end this chapter by giving the BRST and superconformal coordinate
transformations of all fields we use in the construction of
the Polyakov action on the supertorus and a SRS.

Chapter II is devoted to develop a consistent method to first recover the
(well-defined) Polyakov action on the torus \cite{Laz} and then to construct
its extension to the supertorus (see {\bf II}). This method, based on
the covariantization of derivative operators, and the introduction of
geometrical objects such as affine and projective (super) connections,
enables us to find a Polyakov action which is globally defined. In the second
part of this chapter we use the solution to the super Beltrami equations
constructed in chapter I to write this action as a local functional in the
super Beltrami differential and thereby to compute the stress-energy tensor
and its operator product expansions (OPE) corresponding to induced
supergravity described by this
Polyakov action. This will allow us on the other hand to recover the
corresponding results on the supercomplex plane and the torus (see {\bf III}).

In chapter III we proceed to the generalization of this formalism to a super
Riemann surface of arbitrary genus. Here the task is much more difficult since
one is led to deal with complicated and poorly known objects. In particular,
one has to find a way to cope with singular superfields. This mainly consists
in choosing a determination of these fields by cutting the SRS around
singular points. Accordingly, we will
show how some of these difficulties have been circumvented ending up with the
effective action for the $N=1$ $2D-$induced conformal supergravity on a
compact SRS (without boundary) $\hat\Sigma$ of genus $g>1$, as the general
solution of the corresponding superconformal Ward identity. This is
accomplished by defining a new super integration theory on $\hat\Sigma$ which
includes new definition of a coboundary Dolbeault operator and formulation of
the super Stokes theorem and residue calculus in the superfield formalism.
Here integration is performed on a finite domain defined as the direct
product of a domain in the underlying Riemann surface and a ``Grassmann
circle''. The finiteness of this domain is a new result with respect to the
Berezin integration method which uses an infinite domain for the Grassmann
variables. Our integration method allows us to carry all residue calculations
in the superfield formalism without having recourse to the cumbersome method
of components. This is impossible with Berezin's integtration method. Another
crucial ingredient is the notion of polydromic fields studied in chapter I.
The resulting action is shown to be
globally defined (and free of singularities) on $\hat\Sigma$. On the other
hand, this action
has been shown to be only defined up to an arbitrary functional. Furthermore,
we show that this solution can be written in two different forms depending
on whether the fields we start with are single-- or multi--valued. The
solution that starts with single-valued fields is related to the Polyakov
action on the supertorus by restricting all fields from the SRS onto the
supertorus. Moreover, this action
is used to give the most general expression for the stress-energy tensor which
is regular due to a mechanism of compensation of singularities. However, in
order to compute N-point Green functions, defined as the $N^{th}$ order
derivatives of the action with respect to the super Beltrami differential,
we need to find the exact Cauchy kernel on a SRS for solving
the super Beltrami equations thereon, and thus generalize the work of
{\bf III} to this case.

\section{Super Riemann surfaces (SRS)}
It is by now agreed that moduli spaces of SRS are basic objects of superstring
theory\cite{MNP}, mainly because of the possibility
to define a functional on the set of functions, which is independent of the
local structure of the SRS and is invariant under the world sheet supersymmetry
transformations.\\
Teichmuller theory for SRS is mathematically rigorously developed using the
supermanifold theory of Rogers\cite{Rog1}. The basic difference between Rogers'
theory and that of De Witt is that the former allows for a general situation
in which non-trivial topology in the $\theta$ dimensions is possible, while
De Witt topology is trivial in all but the dimension defined by the body $z_0$
of the complex coordinate $z$, see below. Moreover, such a De Witt
supermanifold
theory is more suitable for applications to superstrings\cite{CraBin1} and, in
particular, for a picture of a moving superstring in spacetime. This we will
adopt for our development.\\

\subsection{Geometrical structure}
Consider a $(1\vert 1)-$dimensional complex supermanifold $\hat M$
\footnote{An $(m\vert n)-$dimensional supermanifold is similarly defined by
gluing patches from I$\! \! \!$C$^{m\vert n}$.} which is
obtained by patching together superdomains in the super complex plane
I$\! \! \!$C$^{1\vert 1}$ where the local coordinate charts are
$(U,(z,\theta))$, with $(z,\theta)$ a pair of commuting and anticommuting
complex coordinates \cite{CraBin1,CraBin2,Rog1}.
The underlying manifold of $\hat{M}$ obtained by switching off the
nilpotent elements of $\hat{M}$ is called its body and is denoted by $M$.
The coordinates $(z,\theta)$ take their values
in a Grassmann algebra $I\! \!B_{L}$ with generators
$1,\nu_{1},\nu_{2},\dots ,\nu_{L},$
satisfying $(\nu_{i}\nu_{j}=-\nu_{j}\nu_{i})$ i.e.,
\begin{eqnarray*}
\left \{
\begin{array}{lll}
z &=& z_{0}+z_{ij}\nu_{i}\nu_{j}+\dots	\nonumber \\
\theta &=& \theta_{i}\nu_{i}+\theta_{ijk}\nu_{i}\nu_{j}\nu_{k}+\dots
\end{array}
\right.
\end{eqnarray*}
The coefficients $z_{0},z_{ij},\dots;\theta_{i},\theta_{ijk},\dots $ are
ordinary
complex numbers. $z_{0}$ is called the body of $z$ and $z-z_{0}$ its soul; note
that $\theta$ is pure soul \cite{Rog1,CraBin2}.\\
To make the supermanifold $\hat M$ into a SRS we need the following
additional structures:\\

{\large \bf Super complex structure}\\
This means that the transition functions of $\hat M$ between two charts
$(U,(z,\theta))$ and $(V,(\tilde z,\tilde\theta))$ are complex analytic and
superanalytic, that is, they are of the form
\begin{eqnarray*}
\left \{
\begin{array}{lll}
\tilde z &=& f(z)+\theta\zeta(z) \nonumber \\
\tilde\theta &=& \psi(z)+\theta g(z)
\end{array}
\right.
\end{eqnarray*}
where the component functions $f,\zeta,\psi,g$ have Taylor expansions, e.g.,
$$f(z) = f(z_{0})+(z-z_{0})f^{'}(z_{0})+\dots$$\\
Note that this series terminates because the soul $z-z_{0}$ is nilpotent and
hence
these component functions are uniquely specified by giving their values at
$z_{0}$.\\
However, a super complex structure is not sufficient to define a super
Cauchy-Riemann operator and thereby the action depending on it; for this we
further need a\\

{\large \bf Superconformal structure}\\
This requires that the derivative operator
\begin{equation}\label{chia}
D\equiv D_{\theta} = \partial_{\theta}+\theta\partial_{z},
\hspace{2 cm} D^{2}_{\theta}=\partial_{z}\equiv\partial
\end{equation}
transforms homogeneously. More precisely we have
\begin{equation}\label{chib}
\tilde D = (D\tilde\theta)^{-1}D
\end{equation}
and this consequently imposes the constraint
\begin{equation}\label{chic}
 D_{\theta}\tilde{z}=\tilde{\theta}D_{\theta}\tilde{\theta}
 \end{equation}

or equivalently
\begin{eqnarray*}
\left \{
\begin{array}{lll}
\zeta &=& g\psi\\
g^{2}& =& \partial f+\psi\partial\psi
\end{array}
\right.
\end{eqnarray*}
Finally a general form of the transition functions of a SRS reads,
\begin{eqnarray}\label{chid}
\left \{
\begin{array}{lll}
\tilde z &=& f(z)+\theta\psi \sqrt{\partial f} \\
\tilde\theta &=& \psi(z)+\theta \sqrt {\partial f+\psi\partial\psi}
\end{array}
\right.
\end{eqnarray}
These functions are specified by the two $I\! \!B_{L-1}$-valued
\footnote{We take $(L-1)$ instead of $L$ generators to avoid contradiction
with the Leibniz rule when differentiating with repect to $\theta$\cite{Rog1}.}
analytic functions $f(z_{0})$ and $\psi(z_{0})$, see \cite{CraBin1,RSV}.

We will more precisely consider a compact SRS $\hat\Sigma$ (without boundary)
of genus $g$ with compact Riemann surface $\Sigma$ of genus $g$ with a
particular
spin structure as its body. The charts on $\Sigma$ are the projections on the
$z_{0}$ plane
of the charts on $\hat\Sigma$, and its transition functions are the bodies
$f_{0}(z_{0})$. Thus the transition functions of the tangent bundle over
$\Sigma$
are $f^{'}(z_{0})$, and a spin structure on $\Sigma$ is defined by a choice of
the square root of $f^{'}(z_{0})$. In this setting $\hat\Sigma$ can only be
regarded as a fiber bundle over $\Sigma$ having a vector space as fiber, and
not as a vector bundle since the transition functions may happen to be
nonlinear
in the fiber coordinates\cite{CraBin2}. Unfortunately, not all that we know in
ordinary algebraic geometry carries over onto an arbitrary (compact) SRS.
This is in fact true only in a somewhat trivial case of a split (family of)
SRS,
i.e. for which there are no odd supermoduli parameters ($\psi = 0$, in
(\ref{chid})),
and hence the transition functions for such a SRS become
\begin{eqnarray}\label{chie}
\left \{
\begin{array}{lll}
\tilde z &=& f(z)\\
\tilde\theta &=& \theta \sqrt {\partial f}
\end{array}
\right.
\end{eqnarray}
This is, for instance, the SRS $\hat\Sigma$ obtained from a Riemann surface
$\Sigma$ with an even (non-trivial) spin structure ( see below for the case
$g=1$). On the other hand, the property of splitness (or at least
projectedness)
is necessary to obtain a non-ambiguous rule for integration on a non-compact
supermanifold. However, our integration procedure will be formulated on a
general (but compact) super Riemann surface with De Witt topology, see
chapter III.

In an older approach one treats a SRS as an object representing a
superconformal equivalence
class of 2-dimensional supergravity geometries. Here one starts with a real
$(2\vert 2)-$dimensional manifold $\hat M$, e.g. the one obtained from $M$ with
a spin structure, then introduces a family of frames $\{E_{A}\}$ containing
even as well as odd vectors. Some of these, however, happen to be spurious
owing to
local invariance. Thus one is led to impose gauge conditions so as to end up
with only a graviton and a gravitino of supergravity. This is done in an
indirect
way by introducing still more degrees of freedom in the form of a connection
$\phi_{A}$ on $\hat M$, then one defines the corresponding torsion and
curvature,
and finally imposes conditions on these tensors. Under these conditions and
the ones
they entail via the Bianchi identities, the initial degrees of freedom boil
down to
the minimal graviton--gravitino doublet as required. These torsion constraints
are of two kinds:
the complex torsion constraints implement local equivalence between the almost-
complex structure defined by $E_{A}$ and the standard complex structure of
I$\! \! \!$C$^{1\vert 1}$. Similarly, one gets a local equivalence between the
almost-superconformal structure also defined by $E_{A}$, and the standard flat
superconformal structure of I$\! \! \!$C$^{1\vert 1}$, by imposing
superconformal
torsion constraints\cite{GidNel}. Thus one obtains the two structures
needed to make $\hat M$ into a SRS, as in the previous approach.\\

\subsection{Uniformization theory}

Let us denote by I$\! \! \!$C, $I\! \!P^{1}$ = I$\! \! \!$C$\cup\{\infty\}$ and
$U=\{z\in$ I$\! \! \!$C$; Imz >0\}$ the complex plane, the Riemann sphere and
the
upper half-plane respectively \cite{FarKra}. Then the uniformization theorem
states that
any SRS with metric\footnote{This condition was shown to be superfluous
by Hodgkin\cite{Hodg}. However existence of a metric on $\hat\Sigma$ is
necessary for applications to superstrings.} is $SI\! \!P^{1}$ ( $S$ stands
for super ) or a
quotient of SI$\! \! \!$C or $SU$ by a subgroup $G$ of
\mbox{$SPL(2, $I$\! \! \!$C)},
the group of superconformal automorphisms of $SI\! \!P^{1}$.\\
By imposing analytic conditions on the transition functions (\ref{chid}) at the
point $(z,\theta)=(\infty,0)$, and taking into account the fact that the body
of $f(z_{0})$ must be a conformal automorphism (Mobius transformation) of
$I\! \!P^{1}$, that is
$f_{0}(z_{0})=\frac{a_{0}z_{0}+b_{0}}{c_{0}z_{0}+d_{0}}$, we obtain the
transition functions

\begin{eqnarray}\label{chif}
\left\{
\begin{array}{lll}
\tilde z &=& {\displaystyle\frac{az+b}{cz+d} +
\theta\frac{\gamma z+\delta}{(cz+d)^{2}}} \vspace{0.5cm} \\
\tilde\theta &=& {\displaystyle\frac{\gamma z+\delta}{cz+d} +
\theta\frac{1+\frac{1}{2}\delta\gamma}{cz+d}}
\end{array}
\right.
\end{eqnarray}

$a,b,c,d\in$ I$\! \! \!$C with $ad-bc=1$, while $\gamma$ and $\delta$
are odd Grassmann numbers. These depend on three independent even
parameters in $I\! \!B_{L-1}$ and two odd ones, they generate the group
$SPL(2, $I$\! \! \!$C).\\
Indeed, I$\! \! \!$C, $I\! \!P^{1}$ and $U$ have unique spin structures since
they are simply connected, i.e. their fundamental groups are trivial. Hence
 SI$\! \! \!$C, $SI\! \!P^{1}$ and $SU$ are the unique canonical SRS's over
these Riemann surfaces with De Witt topology.\\
As in the bosonic case, the subgroup $G$ of superconformal automorphisms by
which we take the quotient of these SRS's is isomorphic to their fundamental
group $\pi_{1}(\hat\Sigma)$ which is isomorphic in turn to
$\pi_{1}(\Sigma)$
since, as mentioned previously, $\hat\Sigma$ is a vector-space-fibered fiber
bundle. Therefore, $G$ is discrete since $\pi_{1}(\Sigma)$ is.\\
Thus any SRS is either $SI\! \!P^{1}$ or a quotient of SI$\! \! \!$C or $SU$
by a group $G$
of superconformal automorphisms which acts properly discontinuously on the
body; but no new SRS's can be obtained as quotients of S$I\! \!P^{1}$ since no
Mobius transformation acts properly discontinuously on the body $I\! \!P^{1}$
because this contains the point at $\infty$. This is why a superconformal
automorphism
of SI$\! \! \!$C and $SU$ needs only have a Mobius transformation as its
body while its soul remains unrestricted.\\
As we are considering later on the construction of the Weierstrass
$\zeta-$function and Polyakov action on the
supertorus, it is worthwhile to discuss in more detail the uniformization
construction for
the genus 1 case\cite{CraBin1,FreRab}.\\
The supertorus is then the quotient of SI$\! \! \!$C by a subgroup $G_{1}$
of $SPL(2, $I$\! \! \!$C). Then by requiring of $G_{1}$ to act properly
discontinuously
on I$\! \! \!$C, and to preserve the metric
$ds = \vert dz + \theta d\theta\vert$
on SI$\! \! \!$C we find the conditions $c=0,\gamma=0,a^{2}=1$ and
$b_{0}\neq 0$. Moreover, $G_{1}$ must be abelian as it is isomorphic to the
fundamental
group of the torus, an abelian group with two generators. One next reduces the
generators of $G_{1}$ by conjugation with $SPL(2, $I$\! \! \!$C) elements.
If we
denote by $(a,b,\delta)$ and $(a^{'},b^{'},\delta^{'})$ the two generators of
$G_{1}$ we see that the conditions $a^{2}=1, (a^{'})^{2}=1$ determine four
(three even and one odd) spin structures on the torus which correspond to the
choice
of signs for $a$ and $a^{'}$. Taking all these considerations into account we
can finally write down the generators of $G_{1}$ in different spin structures,
\begin{eqnarray}\label{chig}
\left\{
\begin{array}{lll}
z \rightarrow z + 1 &,& \hspace{1cm} \theta\rightarrow\theta \\
z \rightarrow z + \tau + \theta\delta &,& \hspace{1cm} \theta\rightarrow\theta
+ \delta
\end{array}
\right.
\end{eqnarray}
where $\tau = b$ with $Im\tau \neq 0$, for the odd spin structure $(+,+)$
and
\begin{eqnarray}\label{chih}
\left\{
\begin{array}{lll}
z \rightarrow z + 1 &,& \hspace{1cm}\theta\rightarrow +\theta \\
z \rightarrow z + \tau &,& \hspace{1cm} \theta\rightarrow -\theta
\end{array}
\right.
\end{eqnarray}
for the even spin structure $(+,-)$.\\
The other two even spin structures are obtained by changing the signs in the
transformations of $\theta$ in Eq.(\ref{chih}) from $(+,-)$ to $(-,+)$ or
$(-,-)$.\\
For completeness, we state some results from the uniformization theory on the
super Teichmuller space $ST_{g}(\hat\Sigma)$ of $\hat\Sigma$
\cite{Hodg,CraBin1}.
$ST_{g}(\hat\Sigma)$ is a complex analytic supermanifold of dimension
$(3g-3\vert 2g-2)$ if $g>1$, $(1\vert 1)$ if $g=1$ in the trivial $(+,+)$
underlying spin structure, and $(1\vert 0)$ if $g=1$ in the non-trivial even
spin structures. The body of $ST_{g}(\hat\Sigma)$ is the ordinary Teichmuller
space of
Riemann surfaces with spin structure, this is a $2^{2g}-$sheeted covering of
the Teichmuller
space $T_{g}(\Sigma)$. As was shown by Crane and Rabin \cite{CraBin1}, the
supermodular group which acts by changing the choice of generators of
$\pi_{1}(\hat\Sigma)$ (i.e. a marking of $\hat\Sigma$) is the ordinary modular
group. This reduces the super Teichmuller space to the supermoduli space
${\cal M}_{g}(\hat\Sigma)$, needed in the Polyakov path integral formalism.

\subsection{Super Beltrami  equations : A solution on the supertorus}
A super Riemann surface can further be provided with another superconformal
structure which is represented by a set of isothermal (or projective)
coordinates $(\hat{Z},\hat{\Theta})$\footnote{Henceforth, we will only
consider the holomorphic sector, knowing that everything proceeds along
similar lines
in the antiholomorphic sector (with barred variables)} together with
superconformal transition functions between two charts
$(U,(\hat Z,\hat\Theta))$
and $(V,(\tilde{\hat Z},\tilde{\hat\Theta}))$, i.e.
$D_{\bar{\hat\Theta}}\tilde{\hat Z} = 0,\;
D_{\bar{\hat\Theta}}\tilde{\hat\Theta} = 0,\;
D_{\hat\Theta}\tilde{\hat Z} =
\tilde{\hat\Theta} D_{\hat\Theta}\tilde{\hat\Theta}.\;$
These isothermal coordinates satisfy in the reference structure the
super Beltrami equations (SBE) \cite{CraBin1,Tak,HKMK1,HKMK2},

\begin{eqnarray}\label{chi1a}
\Delta_{\hat\mu} \hat Z &=& -\hat\Theta\Delta_{\hat\mu}\hat\Theta \nonumber\\
\Delta_\nu\hat Z &=&\hat\Theta\Delta_\nu\hat\Theta \nonumber\\
\Delta_\sigma\hat Z &=&\hat\Theta\Delta_\sigma\hat\Theta
\end{eqnarray}
with
$\Delta_{\hat\mu}=\bar\partial -\hat\mu \partial,\quad \Delta_\nu
=\partial_\theta+\nu\partial,\quad
\Delta_\sigma =\bar\partial_\theta +\sigma\partial$,
where $\hat\mu ,\nu$ and $\sigma$ are the super Beltrami differentials
which parametrize the superconformal structure $\{ (\hat Z,\hat\Theta )\}$
and satisfy certain boundary conditions\cite{Tak}.
We will refer to the formalism using these super Beltrami differentials as
the $\hat\mu-$formalism. There is another formalism \footnote
{In the sequel we will use both formalisms alternatively depending
on the technical convenience that each of them might present.} which makes
use of the super Beltrami differentials $H^z_{\bar z}, H^z_\theta,
H^z_{\bar\theta}$ (this will be referred to as the $H-$formalism) and in
which these equations become\cite{DelGie}

\begin{eqnarray}\label{chi1b}
\bar\partial\hat Z+\hat\Theta\bar\partial\hat\Theta &=&
H^z_{\bar z} (\partial\hat Z+\hat\Theta\partial\hat\Theta )  \nonumber \\
D_\theta\hat Z-\hat\Theta D_\theta\hat\Theta &=&
H^z_\theta (\partial\hat Z+\hat\Theta\partial\hat\Theta )   \nonumber\\
\bar D_\theta\hat Z-\hat\Theta\bar D_\theta\hat\Theta &=&
H^z_{\bar\theta} (\partial\hat Z+\hat\Theta\partial\hat\Theta )
\end{eqnarray}

These two sets of equations are related by the
following identification\cite{HKMK1}

\begin{equation}\label{chi1c}
 H^z_{\bar z} =\hat\mu,~~H^z_\theta =\theta -\nu,~~
 H^z_{\bar\theta} = -\sigma +\bar\theta\hat\mu\ .
 \end{equation}
 One should note here that $\hat\mu$ is the supersymmetric extension
of the bosonic Beltrami differential $\mu$\cite{DelGie,AK1,HKMK1}.

Just as in the bosonic case\cite{Laz}, we find that a superconformal
change of coordinates in the SB--structure \footnote{SB stands for super
Beltrami differentials, and SB--structure for the structure defined by
these differentials. Objects that are holomorphic in this structure will be
said to be SB-holomorphic.} $(\hat Z,\hat\Theta )$ induces a local change of
coordinates in the reference structure via a sense-preserving
\footnote{Both the super Jacobian and the Jacobian of the body map are
positive.} superdiffeomorphism $\varphi\in S{\rm Diff}^+(\hat\Sigma )$,
that is, the following diagram commutes

$$\matrix{
(z,\theta ) &\ \ {(\hat\mu ,\nu ,\sigma )\atop\hbox to 40pt{\rightarrowfill}}
 &\quad (\hat Z,\hat\Theta )\hfill\cr  &&\cr
 \varphi\Big\downarrow\quad &\qquad\qquad\qquad
 &\Big\downarrow {\rm superholomorphic}\cr
 &&\qquad\qquad\quad {\rm transformation}\cr
 (\tilde z,\tilde\theta )
 &\ \ {(\tilde{\hat\mu},\tilde\nu ,\tilde\sigma )
 \atop\hbox to 40pt{\rightarrowfill}}
 &\quad (\tilde{\hat Z},\tilde{\hat\Theta})\ .\hfill\cr}$$

This defines a new complex structure on $\hat\Sigma$ parametrized by the new
SB $\tilde{\hat\mu}, \tilde\nu ,\tilde\sigma$.
Then if we denote this superdiffeomorphism by

$$\varphi : (z,\theta )\to (\varphi_b(z,\theta ,\bar z,\bar\theta ),
 \varphi_f(z,\theta ,\bar z,\bar\theta )),$$

we obtain the right action SR$_\varphi$ of
$S{\rm Diff}^+(\hat\Sigma )$ on the
SB as follows\cite{HKMK1,Tak}
\begin{eqnarray}\label{chi1d}
\left.
\begin{array}{llllll}
\tilde{\hat\mu}\equiv SR_\varphi (\hat\mu ) &=&{\displaystyle
{\bar\partial\varphi_b+(\hat\mu\circ\varphi )\bar\partial
\bar\varphi_b+(\nu\circ\varphi )\bar\partial\varphi_f+
(\sigma\circ\varphi )\bar\partial\bar\varphi_f\over
\partial\varphi_b+(\hat\mu\circ\varphi )\partial
\bar\varphi_b+(\nu\circ\varphi )\partial\varphi_f+
(\sigma\circ\varphi )\partial\bar\varphi_f}} \vspace{0.3cm}\\

\tilde\nu\equiv SR_\varphi (\nu ) &=&{\displaystyle{-\partial_\theta
\varphi_b-(\hat\mu\circ\varphi )\partial_\theta
\bar\varphi_b+(\nu\circ\varphi )\partial_\theta\varphi_f+
(\sigma\circ\varphi )\partial_\theta\bar\varphi_f\over
\partial\varphi_b+(\hat\mu\circ\varphi )\partial\bar\varphi_b+
(\nu\circ\varphi )\partial\varphi_f+(\sigma\circ\varphi )
\partial\bar\varphi_f}} \vspace{0.3cm} \\
\tilde\sigma\equiv SR_\varphi (\sigma ) &=&{\displaystyle
{-\bar\partial_\theta\varphi_b-(\hat\mu\circ\varphi )
\bar\partial_\theta\bar\varphi_b+(\nu\circ\varphi )
\bar\partial_\theta\varphi_f+(\sigma\circ\varphi )
\bar\partial_\theta\bar\varphi_f\over
\partial\varphi_b+(\hat\mu\circ\varphi )\partial\bar\varphi_b+
(\nu\circ\varphi )\partial\varphi_f+(\sigma\circ\varphi )
\partial\bar\varphi_f}}
\end{array}
\right.
\end{eqnarray}

This defines a deformation
of the super complex structure defined by $(\hat\mu ,\nu ,\sigma )$ under
the right action of a superdiffeomorphism $\varphi\in S{\rm Diff}^+(\hat
\Sigma )$, i.e.,

$$(\hat\mu ,\nu ,\sigma )\to (\hat\mu^\varphi ,\nu^\varphi ,\sigma^\varphi )$$

In other words, the set $\{ (\hat\mu^\varphi ,\nu^\varphi ,\sigma^\varphi ),
\varphi\in S{\rm Diff}^+(\hat\Sigma )\}$ describes the
$S{\rm Diff}^+(\hat\Sigma )$--orbit of the point $(\hat\mu ,\nu ,\sigma )$ in
the supermoduli space $\hat{\cal M} (\hat\Sigma )$ of $\hat\Sigma$ introduced
earlier.

One should note that the right action SR$_\varphi$ given above
is holomorphic w.r.t. the SB. Moreover, it is straightforward to check that
Eqs.(\ref{chi1d}) allow us to recover those given in\cite{DelGie} through the
identification
(\ref{chi1c}) and that they reduce to the right action of
$\varphi_{0}\in {\rm Diff} (\Sigma )$ (the body of $\varphi$)
on the bosonic Beltrami differential $\mu$ \cite{Laz}.

The Beltrami differentials are viewed as superdifferentials
defined on the upper-half plane and invariant under a subgroup of the
covering group $SPL(2,{\bf R})$ of the super Riemann surface\cite{CraBin1,Mar}.
In our setting these differentials, in particular $\hat\mu$, are
extended to the entire supercomplex plane by defining $\hat\mu = 0$,
$\nu=\theta$ and $\sigma=0$ in the lower-half plane \cite{Mar,CraBin1}.

Now we stress that we will be considering, as discussed in the introduction,
(1,0)-induced
supergravity involving only the field $\hat\mu$. Thus we restrict ourselves to
the following ``gauge choice''\footnote{\normalsize This gauge was conjectured
in \cite{CraBin1} and shown in
\cite{Tak} to be always possible. This is equivalent to
$H^{z}_{\theta}=0$ with no $\bar\theta-$dependence in the $H$-formalism.}

\begin{eqnarray}\label{chi1e}
\left\{
\begin{array}{lll}
\nu &=& \theta \\
\sigma &=& 0
\end{array}
\right.
\end{eqnarray}

The first equation is equivalent \cite{HKMK1} to the condition implying the
existence of superconformal structure, i.e. (\ref{chic}), while the second
one implies that there is no $\bar\theta$-dependence. In this case,
Eqs.(\ref{chi1a}) reduce to
\begin{eqnarray}\label{chi1f}
\Delta_{\hat\mu} \hat Z &=& -\hat\Theta\Delta_{\hat\mu} \hat\Theta\nonumber\\
D\hat Z &=& \hat\Theta D\hat\Theta\ ,
\end{eqnarray}
where now
$\hat Z = \hat Z (z,\theta,\bar z),\hat\Theta
=\hat\Theta(z,\theta,\bar z)$.

The Ward identity (\ref{intrd}) is also the holomorphy equation (in the
$\hat\mu-$structure ) for the
projective connection $\displaystyle{\delta\Gamma\!\over \delta\hat\mu}$,
i.e. the energy-momentum tensor. This is obtained by writing a particular
superprojective connection as
$\hat R=-\partial D\ln(D\hat\Theta)-D\ln(D\hat\Theta)\partial\ln(D\hat
\Theta)$ and then using the decoupled SBE
\begin{eqnarray}\label{chi1g}
\bar\partial\hat Z &=&\hat\mu \partial\hat Z +{1\over 2}D\hat\mu D\hat Z
\nonumber\\
\bar\partial\hat\Theta &=& \hat\mu\partial\hat\Theta +{1\over 2}D\hat\mu
D\hat\Theta
\end{eqnarray}
which were shown in Ref.\cite{Tak} to be equivalent to Eqs.(\ref{chi1f}) in
the gauge (\ref{chi1e}). Furthermore, we
wish to emphasize that holomorphic factorization (which led to the Ward
identity (\ref{intrd})) is in fact defined with respect to
$(\hat Z,\hat\Theta)$ and
($\bar{\hat Z}, \bar{\hat\Theta})$. Moreover, from Eqs.(\ref{chi1g}) we can
see that these coordinates are local functionals in $\hat\mu$ and
$\bar{\hat\mu}$
respectively, thus giving rise to the chiral splitting in Eq.(\ref{intrc}).
Indeed, we showed in {\bf III} that a solution to (\ref{chi1g}) on the
supertorus can be written as
\begin{eqnarray}\label{chi1h}
\hat Z(z,\theta,\bar z)&=&z+\int_{S
T}d\tau\hat\zeta(z-\omega-\theta\phi)[\hat\mu\partial\hat Z+{1\over 2}
D\hat\mu D\hat Z](\omega,\phi,\bar\omega)\\ \nonumber
\hat\Theta(z,\theta,\bar{z})&=&\theta+\int_{ST}
d\tau\hat\zeta(z-\omega-\theta\phi)[\hat\mu\partial\hat\Theta+{1\over
2}D\hat\mu D\hat\Theta](\omega,\phi,\bar\omega)
\end{eqnarray}
where
\begin{equation}\label{chi1i}
\hat\zeta(z-\omega-\theta\phi) = (\phi-\theta)[\zeta(z-\omega)+\phi\theta
p(z-\omega)]=(\phi-\theta)\zeta(z-\omega)
\end{equation}

is  the quasielliptic ``super Weierstrass $\zeta$-function'', i.e.
the $\bar\partial$-Cauchy kernel on the supertorus
$$\bar\partial\hat\zeta(z-\omega-\theta\phi)=-2\pi\delta^{(2)}(z-\omega)
\delta(\theta-\phi).$$
Here $\zeta(z-\omega)$ is the odd quasielliptic Weierstrass $\zeta-$function
and $p(z-\omega)$ is the elliptic Weierstrass function. On the supercomplex
plane we have the Cauchy kernel
\begin{equation}\label{chi1i2}
\frac{\theta-\phi}{z-\omega-\theta\phi}=\frac{\theta-\phi}{z-\omega}
\end{equation}
see discussion in {\bf III}.\\
Let us now discuss the generalization of these objects to a
SRS $\hat\Sigma$ of higher genus $g$. On an ordinary
Riemann surface $\Sigma$, the Cauchy kernel is
$K_\omega(z) =\partial_z\ell n E(z,\omega)$ where $E(z,\omega)$ is the
prime form on $\Sigma$ which has  a simple zero at $z=\omega$, i.e.
$E(z,\omega)\sim(z-\omega)$. On $\hat\Sigma$ we expect the Cauchy kernel
to be of the form $\sim D_1\ell n\hat E(\vec z_1,\vec z_2)$ with
$\vec z_i\equiv(z_i,\theta_i), i=1,2,$ ~where the ``super prime form''
$\hat E$ behaves locally like $(z_1-z_2-\theta_1\theta_2)$.
However, one still needs a more rigorous study of boundary
condition problems in different spin structures on a SRS.

As a last point in this paragraph, we note that in Ref.{\bf I} we have
shown that in the gauge (\ref{chi1e}), the isothermal
coordinates $(\hat Z,\hat\Theta )$ are holomorphic with respect to the
super Beltrami differentials at any point of the SRS $\hat\Sigma$
and in particular at the special choice of origin $\hat P_j$ of this
coordinate system where this property may in general be lost. That is we have
\begin{eqnarray}\label{chi1j}
  \bar{\hat\delta}\hat\Theta_j(\hat P_j) &=& 0 \nonumber\\
  \bar{\hat\delta}\hat Z_j(\hat P_j) &=& 0.
\end{eqnarray}

In the general case (with $\bar\theta$--dependence), the proof of
holomorphy of the projective coordinates proceeds along
similar lines, see Ref.{\bf I} for discussion.
Moreover, the parametrization of the action that
solves the superconformal Ward identity (\ref{intrd}) by the super Beltrami
differentials
provides this action with the important property of holomorphic factorization,
which leads to the corresponding induced Polyakov action. Accordingly, this
property is due to the fact that the coordinates $(\hat{Z},\hat{\Theta})$ are
holomorphic w.r.t. the super Beltrami differentials.
On the other hand, the Ward identity (\ref{intrd}) reflects the non-holomorphy
in the reference structure of the induced action which is a local functional
in $\hat\mu$, a property that is crucial for quantization (e.g. \`a la
Epstein-Glaser \cite{EpGla,Laz}).
Finally, we wish to emphasize that the holomorphy of the isothermal coordinates
$(\hat Z,\hat\Theta)$ w.r.t. SB is compatible with the SB-holomorphy of a
$\demi-$superdifferential $\eta$ on one hand, and w.r.t. the coordinates
$(\hat Z,\hat\Theta)$ on the other.

\subsection{Superdifferentials: Definition}
It is by now well know that the computation of fermionic string amplitudes
in the Polyakov formalism\cite{BMFS,DhPhong1,Pol1} reduces to integration
over a
finite--dimensional superspace, the space of classes of superconformal
manifolds, or equivalently superconformal structures parametrized by
super Beltrami differentials, i.e. the supermoduli space
${\cal M}_{g}(\hat\Sigma)$ introduced above. This parametrization makes the
holomorphic
factorization property manifest at all levels \cite{BBG,KLS,KLT,MKLHL1},
and accordingly the use of holomorphic (super) differentials, among other
geometrical
objects, such as superaffine and superprojective connections,
defined on a super Riemann surface, turn out to be a
powerful means to construct well--defined Lagrangian densities enjoying
this property.
In the bosonic case, these are (1.1)--differential forms \cite{Laz} while in
the supercase these become $(\frac{1}{2},\frac{1}{2})$--superdifferentials.
The properties of a holomorphic differential on a Riemann
surface of genus $g\geq 1$ are well known \cite{FarKra,Gunn,Shaf}, whereas
a complete
study of the same issues in the supercase is still lacking.

Indeed our construction of Polyakov action on a SRS relies crucially on
the holomorphy and monodromy properties of superdifferentials and the
corresponding superconnections and superderivatives. Thus we find it necessary
to discuss (at least) the most important properties of these objects directly
related to our work.\\
For this we consider a compact SRS $\hat\Sigma$ of genus $g$, i.e. with a
compact
underlying Riemann surface $\Sigma$ of the same genus and a particular spin
structure. Next we define a holomorphic line bundle $L$ over $\hat\Sigma$ by
specifying a 1-cocycle $\{ g_{\alpha\beta} \}$ with coefficients in the sheaf
of even holomorphic functions $f$ such that $f(P)$(mod nilpotents)
$\neq 0, \forall P\in\hat\Sigma$. The total space of $L$ can be defined as
the disjoint union
\begin{eqnarray*}
L=\bigcup_{\alpha\in I}(U_{\alpha}\times
\mbox{I}\! \! \!\mbox{C}^{1\vert 1})
\end{eqnarray*}
modulo the equivalence relation
$$(z_{\alpha},s_{\alpha})\sim (z_{\beta},s_{\beta})\longleftrightarrow
(z_{\alpha},s_{\alpha})=(z_{\beta},g_{\alpha\beta}s_{\beta})$$
$\forall (z_{\alpha},s_{\alpha}),(z_{\beta},s_{\beta})\in (U_{\alpha}\cap
U_{\beta})\times $I$\! \! \!$C$^{1\vert 1}$, for
$(U_{\alpha})_{\alpha\in I}$ an
open covering of $\hat\Sigma$.

$g_{\alpha\beta}$ are the transition functions of $L$, which is thus a vector
bundle of rank $(1\vert 0)$ or $(0\vert 1)$ when its sections are even or odd.

Equivalently, $L$ can be characterized by its divisor or as well by its Chern
class $C(L)$. In fact it was shown in \cite{RabTop} that
\begin{equation}\label{chi2a}
C(L)=C(L_{red})
\end{equation}

where $L_{red}$ is the holomorphic line bundle over the reduced SRS
$\hat\Sigma_{red}$, which is just the Riemann surface $\Sigma$ with a spin
structure, i.e. $L_{red}$ is the spin bundle over $\Sigma$.\\
In particular, from Eq.(\ref{chib}) we see that the factor
$F^{-1}=(D_{\theta}\tilde\theta)^{-1}$ is an even transition function of a
line bundle
whose sections are generated by the vector field $D_{\theta}$, i.e. the
tangent bundle $T\hat\Sigma$. To identify the cotangent bundle dual to
$T\hat\Sigma$ we note that the object $d\lambda\equiv (dz\vert d\theta)$
transforms under a superconformal change of coordinates form $(U,(z,\theta))$
to $(V,(\tilde z,\tilde\theta))$ as

\begin{equation}\label{chi2b}
d\tilde\lambda=sdet
\left [\frac{\partial(\tilde z,\tilde\theta)}
{\partial (z,\theta)}\right] d\lambda = F d\lambda
\end{equation}
showing that $d\lambda$ is a section of a fiber bundle with even transition
functions $F=(D_{\theta}\tilde\theta)$
\footnote{$(D_{\theta}\tilde\theta)$(mod nilpotents)=
$\sqrt{\frac{\partial f}{\partial z}}$(mod nilpotents), see eqs.(\ref{chid}),
(\ref{chie})}.
This together with eq.(\ref{chib}) imply that the operator

\begin{equation}\label{chi2c}
\hat\partial \equiv d\lambda\otimes D_{\theta}
\end{equation}
is globally defined; this is the analog of the ordinary Dolbeault operator
$\partial= dz\otimes\partial_{z}$ defined on $\Sigma$. Thus the fiber bundle
dual to $T\hat\Sigma$ is the supercanonical line bundle denoted by
$\hat K$ with the generator $d\lambda$; similarly we define its complex
conjugate $\bar{\hat K}$. $\hat K$ was shown in \cite{HKMK1} to be the square
root of a fiber bundle whose generator is the vierbein field
\begin{equation}\label{chi2d}
E^{\hat Z} = (\partial\hat Z+\hat\Theta\partial\hat\Theta )
(dz+\hat\mu\ d\bar z+\nu d\theta +\sigma d\bar\theta )
\end{equation}

Now we are ready to define a $(\frac{p}{2},\frac{q}{2})-$superdifferential
\footnote{From now on, a comma will separate holomorphic from antiholomorphic
indices while a vertical line will be used to separate even from odd ones.}
 $\Phi$ on $\hat\Sigma$. Here $p,q$ are integers.

 This is a section of the cross fiber bundle
 $\hat K^{\otimes p}\otimes\bar{\hat{K}}^{\otimes q}$ on an open
 set $U\subset\hat\Sigma$. Then if the local cordinates in $U$ are
$(z,\theta)$,
 $\Phi$ reads
\begin{equation}\label{chi2e}
\Phi^{\frac{p}{2},\frac{q}{2}}(z,\theta,\bar z,\bar\theta)=
\phi(z,\theta,\bar z,\bar\theta)
d\lambda^{p}\otimes d\bar\lambda^{q}
\end{equation}
where the coefficients $\{\phi\}$ are smooth functions satisfying the gluing
rule
\begin{equation}\label{chi2f}
 \phi(\tilde{z},\tilde{\theta},\tilde{\bar{z}},\tilde{\bar{\theta}})\;=\;
 (D_{\theta}\tilde{\theta})^{-p}\overline{(D_{\theta}\tilde{\theta})}^{-q}
 \phi(z,\theta,\bar{z},\bar{\theta}).
 \end{equation}
 We also have the identity
\begin{equation}\label{chi2g}
\phi(z,\theta ,\bar z,\bar\theta )=\Lambda^{p/2}\ \bar\Lambda^{q/2}
\psi (\hat Z,\hat\Theta ,\bar{\hat Z},\bar{\hat\Theta}),
\end{equation}
with $\Lambda\equiv\partial\hat Z+\hat\Theta\partial\hat\Theta$,
which expresses the transformation law of the coefficients $\phi$ under
the quasisuperconformal transformation $(z,\theta )\to (\hat Z,\hat
\Theta )$. \\
Let us denote the space of such differentials by $\Omega^{p,q}(U,X)$, where
$X$ is the (monodromy) character to be defined below. In physical terms
$\Phi$ is called a conformal field of weights $(\frac{p}{2},\frac{q}{2})$.

These differentials are more precisely holomorphic in
the complex SB-structure $(\hat Z,\hat\Theta)$.
In the sequel, unless otherwise stated, holomorphy
will always be understood with respect to this structure
\footnote{Holomorphy or superholomorphy will be understood with respect to
the $\mu-$structure $(Z)$ and the SB-structure $(\hat Z,\hat\Theta)$
respectively, whereas the reference structures $(z)$ and $(z,\theta)$ will be
referred to as $0-$structures.}, that is we have the equation
\begin{equation}\label{chi2h}
\bar D_{\hat\Theta} \phi = 0
\end{equation}
If $\Phi$ is holomorphic in the $0$-structure $\{(z,\theta)\}$, i.e.
$\bar D_{\theta}\phi=0$, we will denote it by $\Phi_{0}$, and this notation
will be adopted throughout this thesis.\\
Of particular interest are the $(\frac{1}{2},0)-$superdifferentials
$\Phi^{\frac{1}{2},0}$, which satisfy the holomorphy equation (\ref{chi2h})
which can also be written as
\begin{equation}\label{chi2i}
\begin{array}{l}
\left [ \bar\partial-\hat\mu\partial-\frac{1}{2}(D_{\theta}
\hat\mu)D_{\theta}\right ]
\Phi^{\frac{1}{2},0}\;=\;\frac{1}{2}(\partial\hat\mu)\Phi^{\frac{1}{2},0}
\end{array}
\end{equation}
in the reference structure $\{(z,\theta)\}$; $\partial = \partial_{z}$.\\
At this point we wish to emphasize that eqs.(\ref{chi2f})--(\ref{chi2i}) are
satisfied by every holomorphic $(\frac{p}{2},\frac{q}{2})-$ superdifferential
regardless of its monodromy properties, i.e. with arbitrary character $X$.
However, our subsequent work on a SRS relies crucially on the distinction
between single-valued differentials which generate the space
$\Omega^{p,q}(U,1)\equiv \Omega^{p,q}(U)$,
and multi-valued superdifferentials generating the space $\Omega^{p,q}(U,X)$.
So now we proceed to discuss each of these cases.\\

\subsubsection{Single-valued superdifferentials : Riemann-Roch theorem}

Here (and in the following paragraph ) we specialize to
$(\frac{1}{2},0)-$superdifferentials, though the same discussion
can be addressed to the general case of a $(\frac{p}{2},\frac{q}{2})-$
superdifferential.\\
$\frac{1}{2}-$superdifferentials are sections of the supercanonical line bundle
of rank $(0\vert 1)$, since its generator $d\lambda$ is odd. Then, since its
transition functions are even, these differentials are odd with even
coefficients. In this case, $L=\hat K$ implies $L_{red}=K^{\frac{1}{2}}$,
a spin bundle or the square root of the canonical bundle $K$ over $\Sigma$,
and hence
\begin{equation}\label{chi2j}
C(\hat K) = g-1
\end{equation}
which means that the degree of any section of $\hat K$ is $g-1$, or in
other words
any holomorphic section of $\hat K$ has globally (counting multiplicity)
$g-1$ zeros. Other properties of such differentials can be obtained
from the super Riemann-Roch theorem\cite{Nel,Ninn,RSV}
\begin{equation}\label{chi2k}
dim H^{0}(\hat\Sigma, {\cal O}(L))-dim H^{1}(\hat\Sigma, {\cal O}(L))-
(1\vert 1)C(L)=(1-g\vert 0)
\end{equation}
where ${\cal O}(L)$ denotes the sheaf of holomorphic sections of the
holomorphic
line bundle $L$ over the SRS $\hat\Sigma$. It was shown in \cite{Ninn} that
the
dimension of the space of holomorphic sections of $L$ over a compact
$\hat\Sigma$
is independent of the odd moduli. In the case of a split SRS the theorem
(\ref{chi2k}) follows from the ordinary Riemann-Roch theorem and the identity
\begin{equation}\label{chi2l}
H^{p}(\hat\Sigma, {\cal O}(L)) =
H^{p}(\hat\Sigma_{red}, {\cal O}(L_{red}))\oplus \Pi
H^{p}(\hat\Sigma_{red}, {\cal O}(L_{red}\otimes\hat K_{red}))
\end{equation}
where $\Pi$ is an operator that identifies elements with opposite parities.

For an arbitrary (compact) SRS one uses the index theorem of the operator
$\bar D$, see\cite{RSV}.
As a consequence of this theorem, we compute
$dim H^{0}(\hat\Sigma,{\cal O}(\hat K))$.
For this we put $L=\hat K$ in the theorem (\ref{chi2k}) and then use the
Serre duality\cite{RSV}
\begin{equation}\label{chi2m}
H^{1}(\hat\Sigma, {\cal O}(L))\cong
\Pi H^{0}(\hat\Sigma, {\cal O}(L^{*}\otimes\hat K))^{*}
\end{equation}
where $L^{*}$ is the dual of $L$, to obtain the important result
\begin{equation}\label{chi2n}
dim H^{0}(\hat\Sigma, {\cal O}(\hat K))=(0\vert g)
\end{equation}
which shows that the supercanonical bundle $\hat K$ possesses $g$ odd
holomorphic
sections with even coefficients.\\
To sum up, there are $g$ odd single-valued holomorphic $\frac{1}{2}-$
superdifferentials on a compact SRS $\hat\Sigma$, each of which possesses
globally (counting multiplicity) $g-1$ zeros. These will be henceforth denoted
by $\eta$, and the holomorphic ones in the reference structure by $\eta_{0}$.

Now using this result we can give $\eta$ a local expression in a local
coordinate chart, see {\bf I}. For this let $\hat P_{j}$ denote the $j^{th}$
zero of $\eta$, and $(z_{j},\theta_{j})$ the local coordinates in the chart
$U_{j}$ containing $\hat P_{j}$. Then we locally have
\begin{equation}\label{chi2o}
\eta(\hat P) = \beta(\hat P)(z_{j}(\hat P)-z_{j}(\hat P_{j})-
\theta_{j}(\hat P)\theta_{j}(\hat P_{j}))^{\frac{\alpha_{j}}{2}}
\end{equation}
where $\hat P\in U_{j},\hspace{0.3cm}\alpha_{j}$ is an integer such that
$\sum_{1\leq j \leq N}\frac{\alpha_{j}}{2}= g-1$, and $\beta$ is an even
holomorphic superfunction which does not vanish at $\hat P_{j}$ and which
can be written
in a general form as $$\beta = (D_{\theta}\hat\Theta)G(\hat Z,\hat\Theta)$$
with $G(\hat Z,\hat\Theta)$ an even superfunction such that
$G(\hat P_{j})\neq 0, \forall j=1,2,\ldots,N$. We can give $\eta_{0}$ a similar
local expression around its $k^{th}$ zero $\hat P_{0k}, k=1,2,\ldots,N_{0}$,
of order $\frac{\alpha_{0k}}{2}$ with
$\sum_{1\leq k \leq N_{0}}\frac{\alpha_{0k}}{2}= g-1$,
 using the local coordinates $(z_{k},\theta_{k})$ in a
neighborhood $U_{k}$ containing $\hat P_{0k}$.

Note that in particular, $(D_{\theta}\hat\Theta)$ is the unique ( up to a
holomorphic superfunction )
\footnote{A superholomorphic function F satisfies,
$\bar{D}F = 0$, i.e. $F(z,\theta)=u(z)+\theta v(z)$, and
a meromorphic function on $\hat\Sigma$ is a superfunction which can be
locally written as the ratio of two superholomorphic functions.}
no-where vanishing monodromic $\frac{1}{2}-$superdifferential
\begin{equation}\label{chi2p}
\eta_{T} = (D_{\theta}\hat\Theta)d\lambda
\end{equation}
on the supertorus ($g=1$) with even spin structure, see eqs.(\ref{chi2j}),
(\ref{chi2n}).\\
Similarly, in the $0-$structure on the supertorus, the unique holomorphic
nowhere-vanishing $\frac{1}{2}-$superdifferential $\eta_{0T}$ is given by
\begin{equation}\label{chi2q}
\eta_{0T}= (D_{\theta}\hat\Theta_{0})d\lambda
\end{equation}
locally in an open set $U_{\alpha}$, and satisfies $\bar D\eta_{0T}=0$
\footnote{When no confusion is likely, we will denote a differential $\Phi$ and
its coefficient
$\Phi_{\theta}$ by the same letter in analytic expressions for simplicity},
where $\hat\Theta_{0}=\hat\Theta_{0}(z,\theta)$ is the anticommuting coordinate
in the atlas
$\{(U_{\alpha},(\hat Z_{0\alpha},\hat\Theta_{0\alpha}))\}_{\alpha}$
related to the reference one
$\{(U_{\alpha},(z_{\alpha},\theta_{\alpha}))\}_{\alpha}$
by a superconformal transformation. \\
The differentials $\eta_{T}$ and $\eta_{0T}$ are the building blocks for the
construction of the Polyakov action on the supertorus in much the same way as
$\eta$ and $\eta_{0}$ are for this construction on a SRS.\\

\subsubsection{Multi-valued superdifferentials : Twisted line bundle}

The orbifold formalism has proved more tractable than the Calabi-Yau method
for the compactification problem in superstring theory. This formalism
involves twisted conformal fields, i.e. fields which are
multivalued. Indeed, in discussing interaction on orbifolds one has to consider
multivaluedness of both bosonic and fermionic variables as one goes around the
points on the string world sheet from which spacetime bosons and fermions
respectively are emitted. On the other hand, field twisting in the Polyakov
path integral formalism amounts to inserting spin fields in the
correlation functions of certain vertex operators. For explicit calculations
one
is thus led to choose a determination or sheet of these fields, and show at
the end that physical quantities are independent of the choice of sheet
\cite{HamVaf,Narain,Zu1,Zu2}.\\
Similar treatments were performed to derive the Polyakov action on a (super)
Riemann surface. Indeed Zucchini has found a solution to the conformal
Ward identity (\ref{intrb}) on a Riemann surface of arbitrary genus which
involves
single-valued as well as multivalued holomorphic $1-$differentials \cite{Zu1}.
The consistency of this action and the corresponding surface integrations
relies
crucially on the fact that the multivalued differentials have no zeros. Once
again,
on a SRS we have introduced multivalued $\frac{1}{2}-$superdifferentials to
find a well-defined and a consistent solution to the superconformal Ward
identity
(\ref{intrd}). This motivated us to study in more detail the properties
of such differentials \cite{HKMK1,HK}. To summarize the work done in
\cite{HK} we start with some definitions.

{\bf Definition 1:}
A function element on a compact SRS
$\hat\Sigma$ is a pair $(F,U)$ consisting of an open set $U\subset\hat\Sigma$
together
with a meromorphic superfunction (see footnote 12) $F$ defined on it, i.e.
$F: U\subset\hat\Sigma\longrightarrow SI\! \!P^{1}$.\\
Then, two function elements $(F,U)$ and $(G,V)$ on $\hat\Sigma$ are deemed
equivalent at $\hat P(z,\theta)\in U\cap V$ provided there is an open set
$W\in U\cap V$ containing $\hat P$ such that $F\vert_{W} = G\vert_{W}$.\\
The set of all these classes is called a supergerm of $(F,U)$ at $\hat P$,
which we will denote by ${\cal F}$.\\
Next, we define a path in $\hat\Sigma$ by the map
\begin{eqnarray*}
\gamma: I=[0,1] &\longrightarrow &\hat\Sigma\\
	      t &\longrightarrow &\gamma(t)=\hat P(z,\theta)
\end{eqnarray*}
such that $(\varepsilon^{1\vert 1}\circ\gamma)$ is an ordinary path
\footnote{This means that the body of $\gamma$ is an ordinary path in
$\Sigma$.} in $\Sigma$, where $\varepsilon^{1\vert 1}$ is the body map
defined by
\begin{eqnarray*}
\varepsilon^{1\vert 1}:
I\! \! \!C^{1\vert 1} &\longrightarrow &I\! \! \!C^{1\vert 0}\\
	      (z,\theta) &\longmapsto & \varepsilon^{1\vert 1}(z,\theta)
	      =\varepsilon(z)=z_{0}
\end{eqnarray*}
with $\varepsilon: I\! \!B_{L}\longrightarrow  \;$I$\! \! \!$C.

On the space ${\cal F}$ we define the following projection:
\begin{eqnarray*}
\Pi_{{\cal F}}: {\cal F} &\longrightarrow&\hat\Sigma\\
		(F,\hat P) &\longmapsto \hat P
\end{eqnarray*}
$\Pi_{{\cal F}}^{-1}(\hat P)$ is the germ in ${\cal F}$ above
$\hat P\in\hat\Sigma$.

For our purpose, we consider the special case
in which i) $\Pi_{{\cal F}}$ is surjective and ii) for every path
$\gamma: I\longrightarrow \hat\Sigma$ and every $F\in {\cal F}$ with
$\Pi_{{\cal F}}(F)=\gamma(0)$, there exists a unique path
$\tilde{\gamma} : I\longrightarrow{\cal F}$ in ${\cal F}$ satisfying
$\tilde\gamma(0)=F$, and $\Pi_{{\cal F}}\circ\tilde\gamma=\gamma$. The path
$\tilde\gamma$ is called the {\em analytic continuation} of $\tilde\gamma(0)$
along the path $\gamma$ in $\hat\Sigma$.

Now we define the character $X$ introduced above. For this we consider a
universal covering of $\hat\Sigma$, i.e. a local superdiffeomorphism
$$\pi : \tilde{\hat\Sigma}\longrightarrow\hat\Sigma$$
which satisfies $\pi\circ\phi=\pi,
\forall\phi\in\mbox{SDiff}^{+}(\tilde{\hat\Sigma})\equiv\Gamma$, the group of
sense-preserving (i.e. for which both the super Jacobian and the Jacobian of
the body map
are positive) superdiffeomorphisms of $\tilde{\hat\Sigma}$; it is called the
universal covering group of $\pi$ and is isomorphic to $\pi_{1}(\hat\Sigma)$.
This implies
$$\hat\Sigma = \tilde{\hat\Sigma}/\Gamma.$$
Here $\tilde{\hat\Sigma}$
is a simply connected SRS, i.e. one with a trivial fundamental group.\\

Thus a holomorphic differential on $\hat\Sigma$ can be regarded as a
holomorphic
differential on $\tilde{\hat\Sigma}$ which is invariant under $\Gamma$. Let us
denote by $\Omega^{p,q}(U)$ and $\Omega^{p,q}(\tilde U)$ the space of
differentials defined on $U$ and $\tilde U$ respectively; $\Gamma$ acts on
$\Omega^{p,q}(\tilde U)$ via the pullback.

Now a character ( or multiplier) $X$ on $U\subset\hat\Sigma$ is a map that
associates
to any element $\phi\in\Gamma$ a no-where vanishing element $X(\phi)$ of
$\Omega^{0,0}(\tilde U)$ such that $X(\phi_{1}\circ\phi_{2})= X(\phi_{1})
\cdot X(\phi_{2}),$ with the normalization
$X(\phi_{1}^{-1}\circ\phi_{2}^{-1}\circ\phi_{1}\circ\phi_{2})=1$, for all
$\phi_{1}, \phi_{2}\in\Gamma$.

{\bf Definition 2:}
A (multiplicative) multivalued superfunction belonguing to a character $X$
is a collection ${\cal F}$ of function elements $(F,U)$ on $\hat\Sigma$, with
the following properties:

i) Given two elements $(F_{1},U_{1})$ and $(F_{2},U_{2})$ in ${\cal F}$, then
$(F_{2},U_{2})$ can be obtained by analytic continuation of $(F_{1},U_{1})$
along some curve $\gamma$ on $\hat\Sigma$ and

ii) the continuation of a function element $(F,U)$ in ${\cal F}$ along the
closed curve $\gamma, (\gamma(0)=\gamma(1))$
\footnote{$(\varepsilon^{1\vert 1}\circ\gamma)$ is a closed path in $\Sigma$.
This correspondence establishes the isomorphism between the two fundamental
groups $\pi_{1}(\hat\Sigma)$ and $\pi_{1}(\Sigma)$.} leads to the funtion
element $(X(\gamma)F,U)$ in ${\cal F}$.\\
In general, a polydromic conformal field $\Phi$ of weights $(p/2,q/2)$ with
character $X$ on $U\subset\hat\Sigma$ is an element in $\Omega^{p,q}(\tilde U)$
such that
\begin{equation}\label{chi2r}
\phi^{\ast}\Phi = X(\phi)\Phi.
\end{equation}
$\phi^{\ast}$ denotes the pullback action of the superdiffeomorphism $\phi$
on $\Phi$ \cite{Nel}.\\
For monodromic (single-valued) fields, $X(\phi)=1$ for all
$\phi\in\mbox{SDiff}^{+}(\tilde{\hat\Sigma})$.

If for all $\gamma\in\pi_{1}(\hat\Sigma), \, X(\gamma)$ is constant, then $X$
is the homomorphism\footnote{$I\! \! \!\mbox{C}^{1\vert 1\ast}=
I\! \! \!\mbox{C}^{1\vert 1}-\{(0,0)\}$.}
$$X : \pi_{1}(\hat\Sigma)\longrightarrow I\! \! \!\mbox{C}^{1\vert 1\ast},$$
that is $X\in Hom(\pi_{1}(\hat\Sigma),I\! \! \!\mbox{C}^{1\vert 1\ast})
\cong H^{1}(\hat\Sigma,I\! \! \!\mbox{C}^{1\vert 1\ast})$. This means that
$X$ is a $1-$cocycle on $\hat\Sigma$. In fact $X$ is a homomorphism from
$H_{1}(\hat\Sigma) = \pi_{1}(\hat\Sigma)/[\pi_{1},\pi_{1}]$ to
I$\! \! \!\mbox{C}^{1\vert 1\ast}$, for the latter is commutative with the
multiplication,~$(z_1,\theta_1)\ast(z_2,\theta_2)=
(z_{1}z_{2},z_{1}\theta_2+z_{2}\theta_1)$. The cocycle
condition is guaranteed by the equivalence relation defining the supergerms
belonging to the character $X$. Thus $X$ defines a flat line bundle on
$\hat\Sigma$ called {\em twist line bundle}, and denoted $\xi_{X}$. Thus a
polydromic conformal field
$\Psi$ of weights $(p/2,q/2)$ with multiplier $X$ on $U$ as defined by
eq.(\ref{chi2r}) can be viewed as a smooth section of the twisted line bundle
$\xi_{X}\otimes\hat K^{\otimes p}\otimes\bar{\hat{K}}^{\otimes q}$; the space
of these sections is what we denoted previously by $\Omega^{p,q}(U,X)$.

It is of special interest for our purpose to discuss in more detail the
case of polydromic $\frac{1}{2}-$superdifferentials, i.e. fields of weights
$p=1,\, q=0$, which will be denoted by $\Psi$ or $\Psi_{0}$ when they are
holomorphic in the isothermal or reference structure respectively. These are
sections of the twisted line bundle $\xi_{X}\otimes\hat K$, and thus can,
in general, be written as
$$\Psi = \Psi_{\theta}d\lambda$$
where the coefficient $\Psi_{\theta}$ is even and transforms according to
(\ref{chi2f}) with $p=1,\, q=0$; most importantly one should note that
$\Psi_{\theta}$ is nowhere vanishing. Therefore one can write $\Psi_{\theta}$
locally as\cite{HK,HKMK1}
\begin{equation}\label{chi2s}
\Psi_{\theta}=D_{\theta}\hat{\Theta}h(\hat{Z},\hat\Theta)
\end{equation}
where $h(\hat{Z},\hat\Theta)$ is a multivalued even function belonging to the
character
$X$, which is no-where vanishing\footnote{Since $\xi$ is a flat line
bundle, i.e. the degree of any of its sections is zero.}. The nowhere
vanishing factor ($D_{\theta}\hat{\Theta}$) does make $\Psi$ transform
as stated above, see eq.(\ref{chi2p}). Similarly in the $0-$structure we have
\begin{equation}\label{chi2s2}
\Psi_{0\theta}=D_{\theta}\hat{\Theta_{0}}h_{0}(\hat Z,\hat\Theta),
\end{equation}
$h_{0}$ is likewise an even nowhere vanishing multivalued function.

As emphasized in the beginning of this section, one deals with polydromic
fields by choosing a determination sheet on which these become single-valued.
Accordingly, we define a fundamental domain $\hat D$ of $\pi$ as a simply
connected subset of $\tilde{\hat\Sigma}$ whose underlying domain is the
fundamental domain $D$ of $\Sigma$, with the same properties as $D$
\cite{FreRab}.
$\pi(\hat D)$ is a dissection of $\hat\Sigma$ and $\hat D$ is a $4g-$sided
polygon in $\tilde{\hat\Sigma}$; the vertices of $\hat D$ are mapped by $\pi$
into one and the same point of $\hat\Sigma$, and its sides are mapped into
a set of closed loops based at that point and generate the group
$\pi_{1}(\hat\Sigma)$. This yields a marking of the SRS $\hat\Sigma$.

Now given a polydromic field $\Phi\in\Omega^{p,q}(\tilde{U})$, we define its
branch with respect to a fundamental domain $\hat D$ of $\pi$ as its
restriction
to $\hat D$, that is,
$$\Phi_{\hat D}\equiv\Phi\vert_{\hat D}.$$ Two such branches are related as
follows
\begin{equation}\label{chi2t}
\phi^{*}\Phi_{\hat{D}^{'}}(\hat P)=X(\phi)(\hat{P})\Phi_{\hat D}(\hat P)
\end{equation}
which shows that $\Phi$ picks up a factor when it crosses the boundary of
$\pi(\hat{D})$.
This means that a field which is single-valued on $\tilde{\hat\Sigma}$ with
the boundary conditions (\ref{chi2r}) becomes multivalued on $\hat\Sigma$.

In subsection 3.4.1 below, we will extend the study of polydromic fields by
defining the corresponding branch-cutting of a SRS with De Witt topology.

\subsection{Connections and covariant derivatives}
Using superdifferentials, we can construct other objects which are of crucial
importance for constructing globally defined ( and in compact form ) actions
on a super Riemann surface. More precisely, we can construct out of a
$\frac{1}{2}-$superdifferential objects such as superaffine connections,
superquadratic differentials, covariant derivatives and superprojective
connections.

\subsubsection{Affine superconnections and quadratic superdifferentials}

A superaffine connection $\Xi$ is a collection $\{\Xi_{\theta}\}$ of
superholomorphic functions $\Xi_{\theta}$ in the SB-structure, i.e. satisfying
$D_{\bar{\hat\Theta}}\Xi_{\hat\Theta}=0$, transforming under a superconformal
change of coordinates as follows
\begin{equation}\label{chi3a}
\Xi_{\tilde\theta}=\exp{(\varpi)}(\Xi_{\theta}-D\varpi)
\end{equation}
in $(U,(z,\theta))\cap(V,(\tilde z,\tilde\theta))$, where
$\exp{(-\varpi)}=D\tilde\theta$, with $(D\varpi)^{2}=0$.

Now given a $\frac{1}{2}-$superdifferential $\Phi^{\frac{1}{2},0}$ ( $\eta$ or
$\Psi$ ) we can build a superaffine connection that satisfies this definition.
Explicitly we have,
\begin{equation}\label{chi3b}
\Xi = -D\log{(\Phi^{\frac{1}{2},0})}
\end{equation}
If we consider a single-valued $\frac{1}{2}-$superdifferential $\eta$, the
corresponding superaffine connection, which we will denote by $\zeta_{\theta}$,
possesses logarithmic singularities on a SRS, since $\eta_{\theta}$ has
$g-1$ zeros thereon. But on the supertorus ($g=1$), $\zeta_{\theta}$ is
non-singular.\\
The affine superconnection associated with a multi-valued differential
$\Phi^{\frac{1}{2},0}=\Psi$ will be denoted\footnote{All
subscripts will be dropped later on knowing that analytic expressions involve
the coefficients of the fields and not the fields themselves.} $\chi_{\theta}$.
$\chi$ has no singularities since $\Psi$ is assumed to be nowhere vanishing.\\
Likewise, in the $0-$structure we can define the affine superconnection
$\Xi_{0}$ by
\begin{equation}\label{chi3c}
\Xi_{0}= -D\log{(\Phi_{0}^{\frac{1}{2},0})}
\end{equation}
which satisfies $D_{\bar\theta}\Xi_{0}=0$. For $\Phi_{0}^{\frac{1}{2},0}
=\eta_{0}, \Psi_{0}$, the connection $\Xi_{0}$ will be denoted by $\zeta_{0}$
or $\chi_{0}$ respectively.
The singular properties of these are the same as those of $\zeta$ and $\chi$,
and we will not repeat them here.

In the bosonic case we can build a quadratic differential $Q$ from an abelian
differential $\omega$ and an affine connection $\Gamma$ as follows
$$Q_{\alpha}=\frac{d\Gamma_{\alpha}}{d z_{\alpha}}+
\Gamma_{\alpha}\omega_{\alpha}$$
in $(U_{\alpha},z_{\alpha}), ~U_{\alpha}\subset\Sigma$. In the super case we
have found that the
generalization of this formula yields a superquadratic ( $\frac{3}{2}-$ )
differential $\hat{Q}=\{\hat{Q}_{z\theta}\}$ in terms of a
$\frac{1}{2}-$superdifferential $\Phi^{\frac{1}{2},0}$ and an affine
superconnection $\Xi$. This reads
\begin{equation}\label{chi3d}
\hat{Q}_{z\theta} = \partial\Phi^{\frac{1}{2},0} + \Phi^{\frac{1}{2},0}D\Xi
+ \Xi D\Phi^{\frac{1}{2},0}
\end{equation}
Indeed this differential transforms according to
$$\hat{Q}_{\tilde{z}\tilde{\theta}}=\exp{(3\varpi)}\hat{Q}_{z\theta}$$
The space of superquadratic differentials is dual to the space of super
Beltrami differentials and hence it is the cotangent space to the super
Teichmuller space $ST_{g}$, its dimension is $(3g-3\vert 2g-2)$.
$\hat{Q}=\hat{Q}_{z\theta}(dz\vert d\theta)$ is one of the $(2g-2)$ odd
quadratic differentials.

\subsubsection{Projective superconnections}
These are collections of superholomorphic functions $\{ R^{\Xi}_{z\theta}\}$
i.e. $D_{\bar{\hat\Theta}}R^{\Xi}_{\hat{Z}\hat\Theta}
= 0$, with the following gluing rule

\begin{equation}\label{chi3e}
R^{\Xi}_{\tilde{z}\tilde\theta} = \exp{(3\varpi)}(R^{\Xi}_{z\theta} -
S(\tilde{z},\tilde\theta;z,\theta))
\end{equation}
where $S(\tilde{z},\tilde\theta;z,\theta)$ is the super Schwarzian derivative
$$S(\tilde{z},\tilde\theta;z,\theta) = -e^{-\varpi}\partial D e^{\varpi}
= \frac{1}{2} \theta S(\tilde{z},z),$$
and $S(\tilde{z},z)$ is the Schwarzian derivative.\\
Such superconnection $R^{\Xi}$ can be obtained from an affine superconnection
$\Xi$ through
\begin{equation}\label{chi3f}
R^{\Xi} = -\partial \Xi - \Xi D\Xi
\end{equation}
where $\Xi$ can be either $\zeta$ or $\chi$. The same construction holds in the
$0-$structure using the objects $\zeta_{0}$ and $\chi_{0}$.

Here we wish to note that just as in the bosonic case, the difference of two
affine superconnections is a superabelian ( $\frac{1}{2}-$ ) differential and
that the difference of two projective superconnections ( or
quasi-superquadratic differentials ) is a superquadratic differential.

\subsection{Covariant operators}

To an affine connection $\Xi$ one associates a covariant derivative
$\nabla_{\Xi}$ by defining its action on superfields of conformal weight
$p$ corresponding to the sector $(z,\theta)$\footnote{ For instance
$p(H^{z}_{\bar\theta})=p(\hat\mu)=-1, p(D)=\frac{1}{2}, etc.$} by
\begin{equation}\label{chi3g}
\nabla_{\Xi} = D + 2 p \Xi
\end{equation}
We also define the action of $\nabla_{\Xi}$ on other affine superconnections
$\Xi^{'}$ ($p(\Xi^{'})=\frac{1}{2}$) different from $\Xi$ by
\begin{equation}\label{chi3h}
\nabla_{\Xi}\Xi^{'} = D\Xi^{'} + \Xi~\Xi^{'}
\end{equation}
Furthermore, we define the useful operator
\begin{equation}\label{chi3i}
\Delta_{\Xi} \equiv (\nabla_{\Xi})^{2} = \partial + 2p D\Xi + \Xi D
\end{equation}
which acts on fields of conformal weight $p$. It is important to note that
this operator yields globally defined expressions when it acts on fields of
conformal weight $p=-1$. This observation will be used frequently in the
construction of the well-defined Polyakov action on SRS.

\subsection{BRST and superconformal transformations}
Here we gather, for the sake of completeness, the BRST transformation laws and
the superconformal coordinate change of all fields we will be using in the
following chapters.\\

{\bf Coordinate transformations}
\vspace{0.3cm}

In the sequel we will see ( detailed proofs are given in papers {\bf II} and
{\bf IV} ) that the Polyakov action is globally defined, which means that
it is invariant under a conformal change of coordinates from
$(U,(z,\theta))$ to $(V,(\tilde{z},\tilde{\theta}))$, where $U$ and $V$ are
open subsets of the SRS $\hat\Sigma$. The transformation laws of an affine and
projective superconnections were given in in Eqs. (\ref{chi3a}), (\ref{chi3e})
respectively. In addition we need the following transformation laws

\begin{eqnarray}\label{chi4a}
\tilde{H}&=&\exp{(\bar{\varpi})}\exp{(-2\varpi)}H \nonumber \\
\tilde{\hat\mu}&=&\exp{(2\bar{\varpi})}\exp{(2\varpi)}\hat\mu \nonumber \\
\tilde{\partial}&=&\exp{(2\varpi)}(\partial+D\varpi D) \nonumber \\
\tilde{D}&=&\exp{(\varpi)}D\nonumber \\
\tilde{C}&=&\exp{(-2\varpi)}C
\end{eqnarray}
together with $(D_{\theta}\varpi)^{2}=0$. \\
For the measure

\begin{equation}\label{chi4b}
d^{2}\lambda\equiv\frac{d\lambda\wedge d\bar{\lambda}}{2i}
\end{equation}

and superdifferentials we will use the gluing rules (\ref{chi2b}) and
(\ref{chi2f}) respectively.\\

{\bf BRST transformations}
\vspace{0.3cm}

Since the BRST operator $s$ does not feel the ``Grassmann charge'', i.e.
$deg(s)=0$, the corresponding Leibniz rule that we use throughout our
calculations reads
$$s(\phi\wedge\psi)=s\phi\wedge\psi + \phi\wedge s\psi$$
for the Grassmann-algebra-valued functions $\phi, \psi.$ \\
The action of $s$ on a $\frac{1}{2}-$superdifferential $\Phi^{\frac{1}{2},0}$
is given by
\begin{equation}\label{chi4c}
s\Phi^{\frac{1}{2},0}=-\frac{1}{2}D_{\theta}\Delta_{\Xi} C^{z}
\end{equation}
where $\Xi$ and $\Delta_{\Xi}$ are given by Eqs. (\ref{chi3b}),~(\ref{chi3i}),
this shows that $s$ transforms the projective coordinates
$(\hat Z, \hat\Theta)$ while leaving small coordinates invariant.

For $H^{z}_{\bar\theta}$ we have the following variation in the gauge
$H^{z}_{\theta}=0,$
\begin{equation}\label{chi4d}
sH=\bar{D}C+C\partial H-H\partial C+\frac{1}{2}(DH)DC.
\end{equation}
We also have the following useful identities
\begin{equation} \label{chi4d2}
s(\Delk\Hb)=\Db\Delk\C \hspace{1cm}  \hspace{1cm}  s(\Delta_{\chi}\chi)=
 \demi\Delk\D\Delk\C.
\end{equation}
Similar formulas exist for $\hat\mu$.

Moreover, objects that are holomorphic in the reference structure, such as
the superdifferentials $\Phi_{0}$, the superaffine connections
$\Xi_{0}$ and superprojective connections $\hat{R}_{0}$ ( for instance
the one associated with $\Xi_{0}$ ) are inert under the BRST operator, i.e.
$s\phi_{0}=s\Xi_{0}=s\hat{R}_{0}=0$. \\
we further note that the $s$--variation is assumed to commute with
evaluation at a point of the SRS. Finally, taking into account all these
rules and acting by $s$ to the right one checks its nilpotency.

\section{Induced supergravity on the supertorus}
\subsection{Introduction}
In this chapter we proceed to construct the Polyakov action for $N=1$
2D-induced supergravity on the supertorus (ST). Here there is no problem of
singularities since, according to the
Riemann-Roch theorem, there is a unique holomorphic and no-where vanishing
even superdifferential $\eta$ on ST, see details in paragraph 1.4.
Therefore, objects such as $\partial\ln\eta$, the superaffine connections
(\ref{chi3b}) or the superprojective ones (\ref{chi3f}), which are the
building blocks in the construction of a well-defined action are
non-singular. Thus it is possible to construct a globally defined and
singularity free action for $N=1$ supergravity on the supertorus. As another
important and technically simplifying tool, we use the method of
covariantization of differential operators, this will allow us to find
in an economical way global expressions on a SRS, here the ST. \\
We first apply this method to the bosonic case to rederive the action found
by Lazzarini in\cite{Laz} in a different way. This will already be in a form
suitable for generalization onto ST, yielding the sought-for action.\\
Next we proceed to compute the energy-momentum tensor whose external source
is the super Beltrami differential $\hat\mu$, and operator product expansions
using the action mentioned above. For this task we use the solution
(\ref{chi1h})
to the super Beltrami equations (\ref{chi1g}) to write the action as a formal
power series in $\hat\mu$. Already from the 3-point functions we can see that
this Polyakov action resums the iterative series provided by the renormalized
field theory as a solution to the superconformal Ward identity (\ref{intrd});
thus it satisfies the first Polyakov conjecture concerning uniqueness of
renormalization \cite{Piguet,Bec,Laz}.\\
We will first illustrate this method in the case of the supercomplex plane,
this will be later compared with the analogous results on the supertorus.
Here and in the next chapter we will simplify the notation by dropping all
indices, keeping in mind that we are using the coefficients of differentials
and connections instead of the fields themselves.

\subsection{Covariantization of the Polyakov action on the torus}

We start from the Polyakov action on the complex plane (see \cite{Laz} and
references therein)
\begin{equation}\label{chiia}
\Gamma[\mu]=-\frac{k}{2}\int_{\bf C} d^{2}z \;\mu\partial^{2}log\partial Z
\end{equation}
and rewrite it as the sum of two terms after an
integration by parts
\begin{eqnarray*}
\Gamma[\mu]=k\int_{\bf C} d^{2}z \;A_{T},
\end{eqnarray*}
with
\begin{equation}\label{chiib}
A_{T} = -\mu\partial\xi-\frac{1}{2}\xi\partial\mu
\end{equation}
where
\begin{equation}\label{chiic}
\xi=\partial\log{(\partial Z)}
\end{equation}
is a $\mu$-holomorphic affine connection. Here $Z$ is the projective
coordinate of the complex $\mu-$structure on a Riemann surface $\Sigma$,
which satisfies the Beltrami equation
\begin{equation}\label{chiic2}
(\bar\partial-\mu\partial)Z=0.
\end{equation}
The covariant derivative $\nabla_{\xi}$ associated with $\xi$ reads
\cite{BFIZ,Gie}
\begin{equation}\label{chiid}
\nabla_{\xi}=\partial-p\xi,
\end{equation}
where $p$ is the conformal weight (relative to the $z-$index) of the tensor
on which $\nabla_\xi$ is applied. The action of $\nabla_{\xi}$ on $\xi$
itself is defined as $R_{\xi}\equiv\nabla_{\xi}\xi=\partial\xi-\demi\xi^{2}$,
this is the projective connection associated to $\xi$. \\
Each term in the r.h.s. of Eq.(\ref{chiib}) becomes separately globally
defined on the torus in two steps.
First these terms are covariantized by replacing $\partial$ by $\nabla$ and
the derivative $\partial\xi$ by $\nabla_{\xi}\xi$ to get
\begin{equation}\label{chiie}
A_{T}=\mu(-R_{\xi})\;-\;\frac{1}{2}\xi\nabla\mu.
\end{equation}
Then knowing that the difference of two affine (projective) connections
is an abelian (a quadratic) differential, we introduce standard affine and
projective connections $\xi_{0}$ and $R_{0}$, that are holomorphic in
the reference structure and satisfy ~~$s\xi_{0}=sR_{0}=0$. Note that $\xi_{0}$
is a non-fundamental field in the theory, although it parametrizes
the corresponding conformal model. This arbitrariness is somewhat due to the
freedom we have to choose the coordinate $Z_{0}$ in the reference structure
which is related to $z$ by a conformal transformation i.e.,
$\bar\partial Z_{0}=0$, in terms of which $\xi_{0}$ can be written as
$\xi_{0}=\partial\ln\partial Z_{0}$. \\
This yields the globally defined integrand
\begin{equation}\label{chiif}
A_{T}=\mu(R_{0}-R_{\xi})\;-\;\frac{1}{2}(\xi-\xi_{0})\nabla\mu,
\end{equation}
or explicitly
\begin{eqnarray*}
A_{T}=\mu(R_{0}-\partial\xi+\frac{1}{2}\xi^{2})\;-\;
\frac{1}{2}(\xi-\xi_{0})(\partial+\xi)\mu
\end{eqnarray*}
and then
\begin{equation}\label{chiif2}
\Gamma_{T}[\mu,R_{0}]=\frac{k}{2}\int_{T}d^{2}z
\left[
\mu(R_{0}-\partial\xi-\demi\xi^{2})-\demi(\xi-\xi_{0})(\partial+\xi)\mu
\right]
\end{equation}
is the final expression for the Polyakov action on the torus
found by Lazzarini \cite{Laz};
which yields under the action of the BRST operator the globally defined
(integrated) anomaly, (see \cite{Laz})

\begin{equation}\label{chiig}
{\cal A}(c;\mu;R_{0})=\frac{1}{2}(c\partial^{3}\mu-\mu\partial^{3}c)+
R_{0}(c\partial\mu-\mu\partial c),
\end{equation}

where $c$ is the ghost parametrizing ordinary diffeomorphisms.\\
Here we can see that the introduction of $R_{0}$ is in fact needed
to get a well-defined anomaly and thereby a well-defined BRST cohomology.
$R_{0}$ together with the Beltrami differential $\mu$ defines the
background geometry on which the model represented by
$\Gamma_{T}[\mu,R_{0}]$ depends.\\

We will see in chapter III that the action (\ref{chiif2}) can be related to
the analogous one on a Riemann surface found by Zucchini \cite{Zu1} by an
appropriate restriction of fields thereon.

\subsection{The globally defined Polyakov action on the supertorus}

Here we come to the construction of a solution to the superconformal Ward
identity (\ref{intre}) on the supertorus as sketched in the
introduction to this chapter, thus generalizing the work in the
previous section. In this section we deal with $(1,0)-$induced
supergravity, and accordingly we adopt the appropriate gauge (\ref{chi1e}).
As noted earlier this is equivalent to $H^{z}_{\theta}=0$ with no
$\bar\theta-$dependence.

The superspace generalization of Polyakov's chiral gauge action
\cite{Pol2} was first constructed by Grundberg and Nakayama \cite{GrunNak} on
the supercomplex plane. This
can be rewritten in the following compact form\cite{DelGie}
\begin{equation}\label{chiih}
\Gamma_{SC}[\hat\mu]=\kappa\int_{\bf SC} d\tau\partial\zeta \hat\mu,
\end{equation}
where $\zeta$ (in fact $\zeta_{\theta}$) is the coefficient of
the superaffine
connection defined in eq.(\ref{chi3b}).
The BRST variation of this functional yields the chiral anomaly in
Eq.(\ref{intre}), exhibiting the non-invariance of the Polyakov action under
a subgroup \footnote {This is defined by the gauge (\ref{chi1e}).} of the
superdiffeomorphism group $SDiff_{0}(\Sigma)$.\\
Here again we aim at transforming this action into a globally defined action on
the supertorus following the procedure sketched in the preceeding section.
Accordingly, we first write the action (\ref{chiih}) as the sum of two terms
modulo an integration by parts, that is

\begin{equation}\label{chiii}
 \Gamma_{ST}[\hat\mu]=\frac{\kappa}{2}\int_{ST}d\tau
    \left [ 4(\partial\zeta)\hat\mu + 2\zeta \partial \hat\mu \right ].
\end{equation}
Then we replace the partial derivative $\partial \hat\mu$ by the covariant
one\footnote{$\hat\mu$ has conformal weigths $(-1,1)$, see (\ref{chi4a}).},
i.e., $\Delta_{\zeta}\hat\mu=(\partial-2D\zeta+\zeta D)\hat\mu$,
according to (\ref{chi3i}), and the derivative of the superaffine
 connection $\partial\zeta$ by the superprojective connection
$\hat R_{\zeta}= -\partial\zeta - \zeta D\zeta.$ Thereby the integrand
density in (\ref{chiii}) becomes
\begin{equation}\label{chiij}
A_{ST}=4\left [-\hat R_{\zeta}\hat\mu+
\frac{1}{2}\zeta\Delta_{\zeta} \hat\mu \right ].
\end{equation}

Here again the first term becomes globally defined when a holomorphic ( in
the reference structure ) superprojective connection $\hat R_{0}$, with
$s\hat R_{0}=0,$ is introduced, since
the resulting expression is the difference of a standard superprojective
connection ( $\hat R_{0})$ and a particular one ( $\hat R_{\zeta}$ ), i.e. a
superquadratic differential.
Next we replace the superaffine connection
$\zeta$ by the superabelian differential $(\zeta-\zeta_{0})$ where
$\zeta_{0}$ is a holomorphic
superaffine connection in the $0-$structure with $s\zeta_{0}=0$. Note that
$\Delta_{\zeta}\hat\mu$ is already globally defined as the covariant form of
$\partial \hat\mu$.\\
Considering all these changes together, we find the integrand of the
globally defined Polyakov action on the supertorus

\begin{equation}\label{chiik}
A_{ST}=4\left [ ( \hat R_{0}-\hat R_{\zeta})\hat\mu+
	  \frac{1}{2}(\zeta-\zeta_{0})\Delta_{\zeta}\hat\mu \right]
\end{equation}
and then the Polyakov action reads
\begin{equation}\label{chiil}
\Gamma_{ST}[\hat\mu,\hat R_{0}]=\kappa\int_{ST}d\tau
\left [2(\hat R_{0}-\hat R_{\zeta})\hat\mu +
(\zeta-\zeta_{0})\Delta_{\zeta}\hat\mu\right]
\end{equation}
Just as in the bosonic case, this action depends on the superconformal
background geometry parametrized by the pair $(\hat\mu,\hat R_{0})$, while
its dependence on $\zeta_{0}$ is not fundamental from the physical viewpoint
since $\zeta_{0}$ does not appear in the anomaly (\ref{intre}) owing to
$s(\zeta_{0}\Delta_{\zeta}\hat\mu) = -\bar\partial(\zeta_{0}\Delta_{\zeta}C)$.
In addition, as we will see later, it does not contribute to the
energy-momentum tensor or its correlation functions either.\\
Now let us sketch the proof that the action (\ref{chiil}) indeed integrates
the Ward identity (\ref{intre}). Accordingly, using the BRST transformation
laws
(\ref{chi4c}), (\ref{chi4d}) and $s\hat R_{0}=s\zeta_{0}=0,$ we obtain
\begin{equation}\label{chiim}
s\Gamma_{ST}[\hat\mu,\hat R_{0}]= \kappa{\cal A}(C,\hat\mu,\hat R_{0})+
\frac{\kappa}{2}\int_{ST} d\tau
\left\{D\phi-\bar{\partial}B_{ST}\right\}
\end{equation}

with
\begin{eqnarray}\label{chiio}
\phi&=&-(\hat R_{0}-\hat R_{\zeta})(CD\hat\mu-\hat\mu DC)+
\Delta_{\zeta}(C\partial \hat\mu-\hat\mu\partial C) \nonumber \\
&+&(\zeta D\zeta-\partial\zeta-\zeta\partial)(CD\hat\mu-\hat\mu DC)+
(\Delta_{\zeta})(DCD\hat\mu)
\end{eqnarray}
and
$$B_{ST}=4\left [ (\hat R_{0}-\hat R_{\zeta})C+\frac{1}{2}
  (\zeta-\zeta_{0})\Delta_{\zeta}C \right ]$$
Note that $B_{ST}$ is identical to $A_{ST}$ in (\ref{chiik})
upon substituting $C$ for $\hat\mu$, and thus it is globally defined since
both $\hat\mu$ and $C$ transform homogeneously under a change of coordinates,
see (\ref{chi4a}).\\
Next we perform a coordinate transformation of $\phi$ according to the rules
(\ref{chi3a}), (\ref{chi3e}) and (\ref{chi4a}) to see that
$\tilde{\phi}\;=\;\exp{(\bar{\varpi})}\phi,$ i.e. $\phi$ is a
$(0,\frac{1}{2})-$superdifferential. Therefore, each term under the integral
in the r.h.s. of
Eq.(\ref{chiim}) is separately globally defined. Moreover, $\phi$ and $B$ are
free from singularities, for they involve only non-singular
and single-valued fields (holomorphic differentials) on the supertorus $(g=1)$
( see chapter I ).
Consequently, the last two integrals in $s\Gamma_{ST}$ vanish since their
integrands are total derivatives of globally defined and single-valued
expressions. This finally shows that the Polyakov action (\ref{chiil})
indeed solves the superconformal Ward identity (\ref{intre}) on the
supertorus.

\subsection{Operator product expansions (OPE) of induced supergravity
on the supertorus}

As mentioned previously, our aim is to compute the energy-momentum
tensor corresponding to the super Beltrami differential $\hat\mu$
and operator product expansions from the induced supergravity action
$\Gamma_{ST}[\hat\mu, \hat R_0]$ on the supertorus.
For this we write $\Gamma_{ST}$ as a perturbative expansion in $\hat\mu$
by substituting the projective coordinates $\hat Z, \hat\Theta$ by their
functional expressions in (\ref{chi1h}).
Let us first consider the case of the supercomplex plane. \\

\subsubsection{OPE on the supercomplex plane}

In this case, to compute the OPE using the action (\ref{chiih}), we
generalized to the supersymmetric case the method based on the Neumann series
to solve the SBE\cite{HKMK2}. This obviously led to the solution given in
(\ref{chi1h}). Let us recall this procedure, see {\bf III}. We first rewrite
the equation for $\hat\Theta$ in eq.(\ref{chi1g}) as
$$\bar\partial\Lambda ={1\over 2} \partial\hat\mu +BD\Lambda$$
where
$B\equiv \hat\mu D+{1\over 2}D\hat\mu$ and
$\; \Lambda =\ell n D\hat\Theta$ satisfies
$\Lambda\vert_{\hat\mu = 0} = 0$, which means that our solution
$\hat\Theta$ on $SC$ is given in (\ref{chi1h}) upon substituting $\hat\zeta$
by the Cauchy kernel in (\ref{chi1i2}).\\
Iteratively one gets the formal series
\begin{equation}\label{chiio2}
\bar\partial \Lambda(z,\theta) =\sum^\infty_{k=1}\lambda_k(z,\theta)
\end{equation}
with

$$\lambda_1={1\over 2} \partial\hat\mu,\quad
\lambda_k=BD\bar\partial^{-1}\lambda_{k-1}, \hspace{1cm} k=1,2,\ldots$$
where $\bar\partial^{-1}F$ is defined by using the Cauchy kernel in
(\ref{chi1i2}), that is
$$(\bar\partial^{-1}F)(z,\theta) =\int_{SC}
d\tau\left({\theta-\phi\over z-\omega-\theta\phi}\right) F(\omega,\phi)$$

To compute 3-point functions we expand $\Lambda$ to the third order in
$\hat\mu$. This becomes, after integrating by parts, as
\begin{eqnarray}\label{chiip}
\Lambda(a_1)&=&{1\over 2}\int_{SC}d\tau_2\partial_1C_{12} \hat\mu(a_2)
\nonumber\\
&&\nonumber\\
&+&{1\over 4} \int_{SC} d\tau_{23} [C_{12}\partial^2_{2}C_{23} +3D_1 C_{12}
\partial_2D_2C_{23}]\hat\mu(a_2)\hat\mu(a_3)+O(\hat\mu^3)
\end{eqnarray}
Here and below we will be using the notation
\begin{eqnarray*}
d\tau_{i_1\dots i_k}&=&d\tau_{i_1\dots }d\tau_{i_k}\quad  \mbox{with}\quad
d\tau_i=dm_i\cdot d \theta_i={dz_i\wedge d\bar z_i\over 2i\pi}d\theta_i,
\nonumber\\
C_{ij}&=&\left({\theta_i-\theta_j\over z_i-z_j-\theta_i\theta_j}\right),\;
\partial_i={\partial\over\partial z_i}, D_i
={\partial\over\partial\theta_i}+\theta_i\partial_i
\end{eqnarray*}
with no summation on $i$ in $D_i$, $a_i=(z_i,\theta_i,\bar z_i)$.\\
In terms of $\Lambda$ the action (\ref{chiih}) reads \cite{GrunNak,DelGie}
\begin{equation}\label{chiiq}
\Gamma_{SC}[\hat\mu] =\kappa\int_{SC} d\tau_1 \partial_1 D_1\Lambda (a_1)
\hat\mu(a_1)
\end{equation}
which yields the energy-momentum tensor
\begin{equation}\label{chiir}
{\cal T}(a_1) ={\delta\Gamma_{SC}\over \delta\hat\mu(a_1)}=-2\kappa
(\partial_1D_1\Lambda(a_1)-D_1\Lambda (a_1)\partial_1\Lambda (a_1))
\end{equation}
Note that this vanishes at $\hat\mu = 0$ identically, which obviously means
that $\hat\mu$ is the exterior source for ${\cal T}$.

Now putting the perturbative series (\ref{chiip}) into (\ref{chiir}) we get
\begin{eqnarray}\label{chiis}
{\cal T}(a_1) &=& \kappa\int_{SC} d\tau_2\partial^2_1D_1C_{12} \hat\mu(a_2)
-\ {\kappa\over 2} \int_{SC}
d\tau_{23}\{\partial_1D_1C_{12}\partial^2_{2} C_{23} \nonumber \\
&&\nonumber\\
&+&3\partial^2_1C_{12}\partial_2 D_2C_{23}+\partial_1
D_1C_{12}\partial^2_1C_{13}\} \hat\mu(a_2)\hat\mu(a_3)+O(\hat\mu^3)
\end{eqnarray}
and thus the 2-point functions read
\begin{equation}\label{chiit}
\langle {\cal T}(a_1) {\cal T}(a_2)\rangle =(-1)^2
{\delta^2\Gamma_{SC}\over
\delta\hat\mu(a_1)\delta\hat\mu(a_2)}\bigg\vert_{\hat\mu= 0} = \kappa
\partial^2_1D_1\left(
{\theta_1-\theta_2\over z_1-z_2-\theta_1\theta_2}\right)
\end{equation}
which satisfy the Ward identity\footnote{$\delta^{(3)}(a_1-a_2)\equiv
\delta^{(2)}(z_1-z_2)\delta(\theta_{1}-\theta_{2})=
\delta^{(2)}(z_1-z_2)(\theta_1-\theta_2).$}
\begin{equation}\label{chiiu}
\bar\partial_1\langle {\cal T}(a_1){\cal T}(a_2)\rangle =-\kappa\pi
\partial^2_1D_1\delta^{(3)}(a_1-a_2)
\end{equation}
Likewise, the 3-point functions are
\begin{eqnarray}\label{chiiv}
\langle {\cal T}(a_1){\cal T}(a_2) {\cal T}(a_3)\rangle &=&
(-1)^3{\delta^3\Gamma_{SC}\over
\delta\hat\mu(a_1)\delta\hat\mu(a_2)\delta\hat\mu(a_3)}\bigg\vert_{\hat\mu =
0} \nonumber\\
&&\nonumber\\
&=& \kappa[\partial_1 D_1 C_{12} \partial^2_2 C_{23}+
3\partial^2_1C_{12}\partial_2
D_2C_{23}+\partial_1D_1C_{12}\partial^2_1C_{13}]
\end{eqnarray}
for which the Ward identity reads

\begin{equation}\label{chiiw}
\bar\partial_1\langle {\cal T}(a_1){\cal T}(a_2){\cal T}(a_3)\rangle=\kappa\pi
\left(\partial_{1}\delta^{(3)}(z_1-z_3)\langle {\cal T}(a_1){\cal T}(a_2)
\rangle
-3\partial_{1}\delta^{(3)}(z_{1}-z_{2})
\langle{\cal T}(a_{1}){\cal T}(a_{3})\rangle\right)
\end{equation}

The computation of $N$-point functions follows similar lines, though it is
so (technically) messy that we have not been able to find compact
formulas. However, at third order in $\hat\mu$ we already see
that our results coincide with those found in \cite{DelGie} (to second
order) using the
Ward identity (\ref{intrd}) for a general iterative functional $\Gamma$. This
means that the Polyakov action (\ref{chiih}) resums the iterative series
$\Gamma$ at least at third order  in $\hat\mu$ and thus satisfies the first
Polyakov conjecture concerning uniqueness of renormalization \cite{Bec}
of the conformal model considered here on the supercomplex plane.

\subsubsection{OPE on the supertorus}
Here we follow the same procedure as in the proceeding section
to compute the energy-momentum
tensor corresponding to the super Beltrami differential $\hat\mu$ and
operator product expansions from the induced supergravity action
$\Gamma_{ST} [\hat\mu,\hat R_0]$ in (\ref{chiil}) on the supertorus.
For this we substitute in $\Gamma_{ST}$ the solution to the super Beltrami
equations (\ref{chi1h}), and then by explicit calculations we get
\begin{equation}\label{chiix}
{\cal T}(a_1)=2\kappa(\hat{R}_{o}-\hat{R_{\zeta}})
\end{equation}
In fact we found that $\delta\Gamma_{ST}=2\kappa(\hat R_{0}-\hat R_{\zeta})
\delta\hat\mu
+\D[\ldots]+\Db[\ldots]$. However, the terms on which $\D$ and $\Db$ act are
well-defined and
non-singular on the supertorus and hence they disappear upon integration.
Now we see that indeed the term containing $\zeta_0$ does not
contribute to the stress-energy tensor of the action (\ref{chiil}).

The procedure that led to the expression of $\Lambda$ in
Eq.(\ref{chiip}) remains the same on the supertorus, but here one uses the
Cauchy kernel in Eq. (\ref{chi1i}) instead of (\ref{chi1i2}), thus we get
\[
\Lambda(a_1) ={1\over
2}\int_{ST_2}d\tau_2\partial_1\hat\zeta(z_{12})\hat\mu(a_2)\]
\[+{1\over 4}
\int_{ST_2}d\tau_{23}\{\hat\zeta(z_{12})\partial^2_2\hat\zeta(z_{23})
+3D_1\hat\zeta(z_{12})D_2\partial_2\hat\zeta(z_{23})\}\hat\mu(a_2)
\hat\mu(a_3)+O(\hat\mu^3)
\]
and then
\begin{eqnarray}\label{chiiy}
{\cal T}(a_1) &=& 2\kappa\hat R_{o} +
\kappa\int_{ST_2}d\tau_{2}\partial^2_1 D_1\hat\zeta(z_{12})\hat\mu(a_2)
-{k\over 2}\int_{ST_2} d\tau_{23}
\{\partial_1D_1\hat\zeta(z_{12})\partial^2_{2}\hat\zeta(z_{23})\nonumber\\
&&\nonumber\\
&+& 3\partial^2_1\hat\zeta(z_{12})D_2\partial_2\hat\zeta(z_{23})
+\partial_1 D_1\hat\zeta(z_{12})\partial^2_1\hat\zeta(z_{13})
\}\hat\mu(a_2)\hat\mu(a_3)
+O(\hat\mu^3)\ .
\end{eqnarray}
which leads to the 2-point functions
\begin{equation}\label{chiiz}
\langle {\cal T}(a_1){\cal T}(a_2)\rangle =
\kappa\partial^2_1D_1\hat\zeta(z_{12})=-k\partial_1\hat p(z_{12})
\end{equation}
and the 3-point functions
\begin{eqnarray}\label{chiia1}
\langle {\cal T}(a_1){\cal T}(a_2){\cal T}(a_3)\rangle &=& \kappa(
\partial_1
D_1\hat\zeta(z_{12})\partial^2_2\hat\zeta(z_{23})+3\partial^2_1
\hat\zeta(z_{12})\partial_2
D_2\hat\zeta(z_{23})\nonumber\\
&+&\partial_1 D_1\hat\zeta(z_{12})\partial^2_1\hat\zeta(z_{13}))
\end{eqnarray}
Note that here we use the fact that the superprojective connection
$\hat R_{0}$ is holomorphic in the reference structure $\{(z,\theta)\}$, and
thus $\displaystyle{\delta \hat R_{0}\over \delta\hat\mu} = 0$.

\noindent
Using the Cauchy kernel (\ref{chi1i}) we find that the Green functions
(\ref{chiiz}) and (\ref{chiia1}) satisfy similar Ward identities as those on
the supercomplex plane.

Now writing (\ref{chiiz}) in components using
${\cal T}(a_i) =S(z_i,\bar z_i)+\theta_i T(z_i,\bar z_i)$ and
Eq.(\ref{chi1i}) we get for the 2-point functions
\begin{eqnarray}\label{chiia2}
\langle T(z_1) T(z_{2})\rangle &=& k\partial^3_1\zeta(z_1-z_2) \nonumber\\
\langle S(z_1)S(z_2)\rangle &=& -k\partial^2_1\zeta(z_1-z_2)
\end{eqnarray}
while
\begin{equation}\label{chiia3}
\langle S(z_1)T(z_1)\rangle = 0 = \langle T(z_1) S(z_1)\rangle\ .
\end{equation}
The same substitutions in the 3-point functions (\ref{chiia1}) yield for the
OPE of three T's
\begin{equation}\label{chiia4}
\langle T(z_1) T(z_2)T(z_3)\rangle = 3k \partial p(z_1-z_2)\partial
p(z_2-z_3)
\end{equation}
Here we stress that the results (\ref{chiia2})--(\ref{chiia4}) coincide with
those
found in the bosonic case \cite{Laz}.

To show that the Polyakov action (\ref{chiil}) resums, just as in the
supercomplex plane, the iterative functional and thus provides unique
renormalization on the supertorus, we have to compare the
results (\ref{chiiz}) and (\ref{chiia1}) with those that can be obtained from
an iterative solution $\Gamma$ of the Ward identity
\[
(\bar\partial-\hat\mu \partial -{1\over 2} D\hat\mu D-{3\over
2}\partial\hat\mu){\delta \Gamma\over \delta\hat\mu} =
\kappa\partial^2 D\hat\mu\
.\]

Accordingly, by using the Cauchy kernel (\ref{chi1i}) we obtain\\

\[{\delta \Gamma\over \delta\hat\mu(a_1)}=
\kappa\int_{ST_2}d\tau_2\partial^2_2
D_2\hat\zeta(z_{12})\hat\mu(a_2)\]
\[+\int_{ST_2}d\tau_2\hat\mu(a_2)\{-{3\over
2}\partial_2\hat\zeta(z_{12})-\hat\zeta(z_{12})\partial_2+{1\over
2}D_2\hat\zeta(z_{12}) D_2\}
{\delta \Gamma\over \delta\hat\mu(a_2)}\]
and thereby
\begin{equation}\label{chiia5}
{\delta^2 \Gamma\over\delta\hat\mu(a_2)\delta\hat\mu(a_1)}
\bigg\vert_{\hat\mu =0}=\kappa\partial^2_2 D_2\hat\zeta (z_{12})+
[-{3\over 2}\partial_2\hat\zeta(z_{12})-\hat\zeta(z_{12})\partial_2+
{1\over 2}D_2\hat\zeta (z_{12})D_2]{\delta \Gamma\over
\delta\hat\mu(a_2)}\bigg\vert_{\hat\mu = 0}
\end{equation}

\noindent
This is the generalization to the supertorus of the O.P.E. for the stress
supertensor $\hat T=\displaystyle{\delta\Gamma\over \delta\hat\mu}$ on the
supercomplex plane first found by Friedan \cite{Fried}.

In components Eq.(\ref{chiia5}) yields the following OPE
\footnote{R.T. stands for regular terms}
\begin{eqnarray}\label{chiib1}
T(z_2,\bar z_2) T(z_1,\bar z_1)
&=&k\partial^3_1\zeta(z_1-z_2)+2\partial_1\zeta(z_1- z_2) T(z_1,\bar
z_1)+\zeta(z_1-z_2)\partial_1 T(z_1,\bar z_1)+R.T.  \nonumber\\
S(z_2,\bar z_2) S(z_1,\bar z_1) &=& k\partial^2_1\zeta(z_1-z_2)
+{1\over2}\zeta(z_1-z_2)T(z_1,\bar z_1)+R.T.  \nonumber\\
T(z_2,\bar z_2)S(z_1,\bar z_1) &=&
{3\over 2}\partial_1\zeta (z_1-z_2)S(z_1,\bar
z_1)+\zeta(z_1-z_2)\partial_1 S(z_1,\bar z_1)+R.T.\nonumber \\
S(z_2,\bar z_2) T(z_1,\bar z_1) &=& {3\over 2}
\partial_1\zeta(z_1-z_2)S(z_1,\bar z_1)+{1\over 2}\zeta (z_1-z_2)\partial_1
S(z_1,\bar z_1)+ R.T.
\end{eqnarray}

Note that on the right hand sides of (\ref{chiib1}) $T$ and $S$ are
components of $\hat T$ at $\hat\mu = 0$. Now if we take $\hat T$ to be
${\cal T}(a_1)$ given in
(\ref{chiix}), which means that $\Gamma$ in this case becomes the Polyakov
action (\ref{chiil}), we easily recover the results in (\ref{chiia2}),
(\ref{chiia3}). Here we use the fact that
$\hat R_0 =S_0+\theta T_0$ does not contribute to the singular terms in the RHS
of (\ref{chiib1}) since it is a holomorphic object.

Finally we note that in Ref.\cite{Laz}, the Schwarzian derivative corresponding
to the Polyakov action on the torus given there was computed in terms of the
solution to the Beltrami equation (\ref{chiic2}) by using the Cauchy kernel
$\zeta(z-\omega)$, but the
corresponding OPE was not given. In principle this could be found by
differentiating the Schwarzian derivative with respect to the Beltrami
differential $\mu$ and bringing out the Schwarzian in the resulting expression.
our formalism, this result is obtained automatically since it corresponds to
one of the OPE's in the bosonic limit, that is the first equation in
Eq. (\ref{chiib1}). From the third equation in (\ref{chiib1}) we see that
$S(z,\theta)$ is the component of the energy-momentum tensor whose
external source is the gravitino field of spin $\frac{3}{2}.$

\subsection{Conclusion}

\indent
Using the solution to super Beltrami equations for the projective
coordinates $(\hat Z,\hat\Theta)$ as functionals of the (unique) super
Beltrami differential $\hat\mu$, we have computed the energy-momentum tensor
and 2-and 3-point functions for the induced supergravity Polyakov action on
the supertorus. From this we recover the results on the supercomplex plane
and the torus. On the other hand, from the Ward identity we have derived the
general OPE on the supertorus and shown that they reduce to those found for
the Polyakov action. This means that this Polyakov action resums, at least to
the third order, the iterative series provided by the renormalized field
theory and thus satisfies the first Polyakov conjecture concerning uniqueness
of renormalization. Actually, explicit calculation showed that this argument
carries over to the fifth order, and we believe that this should hold to all
orders, however we have not been successful in finding compact formulae.

\section{$2D-$induced conformal supergravity on a super Riemann surface (SRS)}
\subsection{Introduction}
On a Riemann surface (RS) life is much more subtle than on the torus
and makes use of more
sophisticated algebraic geometrical techniques. In order to construct a
regular and well-defined effective action for a conformal model, one has to
tackle two related difficulties. The first one is connected with
the singularity problem, for most of fields become singular as differentials
acquire zeros according to Riemann-Roch theorem. The second problem is that
of renormalization ambiguity, which is due to the fact that on a Riemann
surface of genus $g>1$ the conformal Ward operator $(\hat\partial-\mu\partial-
2(\partial\mu))$ possesses $3g-3$ zero modes. On the contrary, on the complex
plane and the torus there are none. This implies that the actions in
(\ref{chiif2}) and (\ref{chiil}) are unique. In contrast, on a Riemann
surface the Polyakov action is only defined up to addition of some functional
\cite{Zu1}. In this reference, the author computed the effective action
for induced conformal gravity on a higher genus Riemann surface which depends
holomorphically on the Beltrami differential, and integrates the
diffeomorphism anomaly thereon. He also showed that this solution can be
written in two different forms involving fields with different monodromy
properties, and he indicated how to go from one solution to
another. However, there was no telling about the physical relevance of either
of these two actions. One possible reason for this failure is, apart from
technical hurdles, the fact that both solutions depend strongly on the
background conformal geometry. Nevertheless, Zucchini has managed to compute
the energy-momentum tensor for the action that starts with single-valued
differentials (see below). The computation of $N-$point functions is still
a challenging problem.

Considering all this, one can imagine that generalizing such issues onto a
super Riemann surface is far from being an obvious task, since in addition
to all the difficulties mentioned above, one is faced with supersymmetric
complications and the lack of appropriate geometrical tools. Though, here
again we have an analog of Riemann-Roch theorem (see chapter I and references
therein) wich tells us that superfields may have zeros on a compact
SRS (without boundary), and thereby objects such as affine and projective
superconnections become singular (see (\ref{chi3b}) and (\ref{chi3f})).
This makes it rather difficult, if at all possible, to expand these
superfields
into their components as there is, in general, no telling which component
exhibits such or such singularity. Consequently, Berezin rules for integrating
Grassmann variables are no longer adequate in such a situation. Instead, we
need to devise a more suitable method to integrate these variables and an
analog of Stokes theorem to convert a surface integral into a line integral,
then use Cauchy theorems to perform the residue calculus.

Accordingly, we have obtained a super analog of Stokes theorem which allows us
to
perform any integral (when only integrable singularities are present)
over a compact SRS by using super Cauchy theorems, without having recourse to
component expansion. Indeed, this was made possible by defining a new
integration
procedure in which we integrate Grassmann variables over some ``Grassmann
circle'' surrounding the singular point instead of the whole Grassmann
algebra, as one does in Berezin's theory.\\
Many technical difficulties have been overcome by means of covariant
derivatives and compact notations which turned out to be really
necessary to carry out the calculations in the superfield formalism.
Another subtle ingredient was the notion of polydromic superfields. This led
us to study the problem of dissecting and branch-cutting a SRS with
De Witt topology; a domain in such a SRS is viewed as the ``product'' of its
underlying domain in the Riemann surface and the ``Grassmann circle''
introduced above. Here we are in particular alluding to domains that enclose
the singularity.

Carrying out the whole task results in obtaining a well-defined and
non-singular expression
for the effective action for $N=1$ ~$2D-$induced conformal supergravity on a
SRS of arbitrary genus $g>1$, which depends holomorphically on the super
Beltrami differential and solves the superconformal Ward identity
(\ref{intre}), here written
using the super Beltrami differential $\Hb$ instead of $\hat\mu$\footnote{
The work on a SRS that we present in this chapter was developed in the
$H-$formalism, however it is a straightforward matter to convert everything
into the $\hat\mu-$formalism, since this only amounts to integrating over
$\bar\theta$ throughout the calculations and using the gauge (\ref{chi1e})}.
Here
again we stress that this solution is only defined up to a BRST-invariant and
holomorphic (with respect to the super Beltrami differential) functional. This
is mainly due to the fact that the Ward operator $[\Db-\Hb\partial
-\frac{3}{2}(\partial\Hb)+\demi(\D\Hb)\D]$ in the gauge $H^{z}_{\theta}=0$,
{}~has
$(3g-3)$ even and $(2g-2)$ odd zero modes on a compact SRS (see paragraphs 1.2
and 1.5 ). Furthermore, we explicitly show that we can write this
solution
in two different forms depending on whether the fields we start with are
single-valued or multivalued. This is of course reminiscent of what happens
in the bosonic case. Unfortunately, we did not have enough time to fathom
the problem of renormalization ambiguity, and so we did not approach the
computation of $N-$point functions on a SRS. Nonetheless,
we give a general expression for the energy-momentum tensor and its OPE
with $\demi-$superdifferentials.
Moreover, we have shown that the second solution which starts
with a set of single-valued differentials can be related to the action
(\ref{chiil}) upon restricting all fields onto the supertorus.\\
In a first section I will try to review as briefly as possible the
corresponding work on a Riemann surface \cite{Zu1}. The corresponding
developments were
the starting point for our generalization onto a SRS. On the other hand
this section will help us to compare both results afterwards.

\subsection{ A Polyakov action on Riemann surfaces}

The effective action that describes the $2D$ quantum gravity in the
light-cone gauge on a Riemann surface (RS) was constructed by Zucchini in
\cite{Zu1}. This can be written as the sum of the following terms

\begin{eqnarray}\label{chiiia}
\Gamma_1[\mu] &=& \int_{\Sigma}d^{2}z
\left[
2(R_0-R_{\xi_{0}})\mu+\partial(\Delta_{\xi_{0}}+\Delta_{\xi})\mu
\right]\nonumber\\
&&\nonumber\\
\Gamma_2[\mu] &=& \sum_{j}\nu_{j}\ln(\Omega/\omega_{0})(P_{j})
+\sum_{k}\nu_{0k}\ln(\omega/\Omega_{0})(P_{0k}),
\end{eqnarray}
$d^{2}z=(d\bar z\wedge dz)/2i\pi$. Let us first explain the notation.\\
The Riemann surface $\Sigma$ is parametrized by a reference complex
structure $\{(z,\bar z)\}$ and an isothermal one represented by the projective
coordinates $(Z,\bar Z)$, obtained from $(z,\bar z)$ by a
quasiconformal transformation parametrized by the Beltrami differential $\mu$
with $\|\mu\|<1$. $Z$ satisfies the Beltrami equation (\ref{chiic2}).
Recall that according to the Riemann-Roch theorem \cite{FarKra,Gunn,Nag,Shaf},
there are $g$ holomorphic (in the projective structure) $1-$differentials,
and that every single-valued differential has precisely $2g-2$ zeros counting
multiplicity on a compact Riemann surface of genus $g$. Let us denote such
differentials generically by $\omega$ or $\omega_0$ when they
are holomorphic in the $\mu-$ or $0-$structure respectively.
Likewise, we denote multivalued differentials by $\Omega$ and $\Omega_0$,
which are, conversely, free of zeros\cite{Zu1}. \\
Now denote by $P_j$ the $j^{th}$ zero of $\omega$ and by $\nu_j$ its
order, $Z_j$ is the nearby projective coordinate normalized so that
$Z_{j}(P_{j})=0$; then we get the singular expansion for $\omega$
$$\ln\omega=\nu_{j}\ln Z_{j}+R.T..$$
Similarly for $\omega_0$ we have
$$\ln\omega_0=\nu_{0k}\ln Z_{0k} +R.T.$$
around the zero $P_{0k}$ of order $\nu_{0k}$. Here the coordinate $Z_{0k}$ is
obtained from $z$ by a conformal transformation, i.e. $\bar\partial Z_{0k}=0$,
it satisfies ~$Z_{0k}(P_{0k})=0$.\\
Using the differentials $\omega$ and $\omega_0$ we define the affine
connections $\xi=\partial\ln\omega,~~\xi_{0}=\partial\ln\omega_0$ to which are
associated the covariant derivatives $\nabla_{\xi}$ and $\nabla_{\xi_{0}}$
according to (\ref{chiid}). Analogous objects can be built out of the
multivalued differentials $\Omega$ and $\Omega_0$. In (\ref{chiiia}) $R_0$ is
a standard projective connection and $R_{\xi_{0}}=\demi(\nabla_{\xi}\xi_0+
\nabla_{\xi_0}\xi)$ is a particular one, see (\ref{chiid}).\\
Note that despite the presence of polydromic fields in $\Gamma_2[\hat\mu]$
above, the
whole expression was shown \cite{Zu1} to be single-valued.  Moreover, if
$\Omega$ had zeros then $\Gamma_2$ would be singular thereat. However, one
could render the whole action $\Gamma =\Gamma_1 + \Gamma_2$ non-singular and
single-valued only if a functional $\Phi(\mu)$ with compensating singularity
and monodromy properties is added. This functional must in addition satisfy
the conditions, $\delta\Phi(\mu)/\delta\bar\mu=0,~~s\Phi(\mu)=0$. This
shows that the whole action $\Gamma$ is, as discussed in the introduction,
only defined up to addition of such a functional. Moreover, the monodromy
behaviour of the fields one starts with has much to do with this ambiguity
problem. Indeed it was shown in Ref.\cite{Zu1} that if instead of $\Gamma_1$
one starts with another action, say $\Gamma_3$, obtained from $\Gamma_1$ by
substituting the multivalued fields $\Omega,\Omega_0$ for the single-valued
ones $\omega, \omega_0$ respectively, one finds a second solution to the
conformal Ward identity (\ref{intrb}). More precisely, one gets
\begin{eqnarray}\label{chiiib}
\Gamma_{3}[\mu] &=& \int_{D}d^{2}z
\left[
2(R_{0} - R_{{\cal F}_{0}})\mu+
\partial(\nabla_{{\cal F}_{0}}+ \nabla_{{\cal F}})\mu
\right] \nonumber\\
\Gamma_{4}[\mu]&=&\frac{1}{2i\pi}\oint_{\partial D(\omega_{0})}
\ln(\Omega_{D}/\omega_{0})d\ln(\omega/\Omega_{0D})\nonumber\\
\Gamma_{5}[\mu] &=& \sum_{k}\nu_{0k}\ln(\omega/\Omega_{0D})(P_{0k}).
\end{eqnarray}
Here $D$ is the fundamental domain in $\Sigma$ that contain all zeros of
$\omega$ and $\omega_0$; ~$\Omega_{D}, \Omega_{0D}$ are restrictions of
$\Omega, \Omega_{0}$ to $D$ respectively; ~$\partial D(\omega_{0})$ is the
countour that encirles only the zeros of $\omega_0$. As it stands,
${\cal F}$ and ${\cal F}_0$ are the affine connections
associated with $\Omega$ and $\Omega_0$ respectively, i.e. ${\cal F}=\partial
\ln\Omega,~~{\cal F}_{0}=\partial\ln\Omega_{0}$; ~~and
$R_{{\cal F}_{0}}= \demi(\nabla_{{\cal F}}{\cal F}_{0}+
\nabla_{{\cal F}_{0}}{\cal F})$. \\
This exhibits once again the ambiguity in defining the Polyakov action. We
believe that this is closely related with its dependence on the geometry of
the Riemann surface on which it is constructed.\\
Finally, we wish to mention that many arguments make us prefer
the first solution (\ref{chiiia}), and let us mention two of them.
In Ref.\cite{Zu1} it was this solution that allowed for the computation of
the energy-momentum tensor; and for us, it is this one that we have managed to
relate to the one found on the torus (\ref{chiif2}) without delving into the
complicated problem of monodromy. For the sake of brevity, we only sketch
how to connect the solution on a RS (\ref{chiiia}) to the one on the torus
(\ref{chiif2}).

\subsection{Relating the Polyakov action on a Riemann surface and on the torus}

Using the notation of the previous section and chapter II, simple algebra
leads to the following identity
\begin{equation}\label{chiiic}
A_{T}=A_{RS} + 2 \partial\bar{\partial}\ln\omega
\end{equation}
which holds on a Riemann surface. $A_{T}$ was given in (\ref{chiif}) and
$A_{RS}$ denotes the integrand density in $\Gamma_{1}$ in (\ref{chiiia}). \\
Then, note that since $\omega_0$ is holomorphic in the
reference
structure, i.e. ~$\bar{\partial}\omega_{0}=0$, the second term in the r.h.s.
of (\ref{chiiic}) can be rewritten
as ~$\partial\bar{\partial}\ln(\omega/\omega_{0})$, which thus becomes
globally defined since $\omega/\omega_{0}$ is now an (invariant) function.
Now if we declare $\omega$ and $\omega_0$ to be defined on the torus, i.e.
on the complex plane and invariant under the covering
group of the torus (a subgroup of $SL(2,I \! \! \!C)$), we see that this term
is a total derivative of a single-valued, well-defined and non-singular
differential on the torus and thus disappears upon integration.
Therefore, the action $\Gamma_{1}$ in (\ref{chiiia}) reduces to the one in
(\ref{chiif2}), while $\Gamma_{2}$ vanishes trivially because
$\nu_{j}=\nu_{0k}=0$ as differentials have no zero on the torus.

Although the torus is always treated separately from the topological
point of view, the possibility to connect the Polyakov action on a
Riemann surface to that on the torus makes us believe that the torus is
not so exceptional, but can rather be related, at least in this context, to
the general case of Riemann surfaces of higher genus.

\subsection{A Polyakov action on a higher genus super Riemann surface}

As mentioned in the introduction, an obvious generalization of (\ref{chiil})
onto a compact super Riemann surface of genus $g>1$ can be inspired from the
action in (\ref{chiiib}) by replacing the geometrical objects appearing in it
by their supersymmetric analogs. Thus we start with the following action
\begin{equation} \label{chiiid}
\Gamma_{a}[\Hb,\hat{R_{0}}]=\frac{k}{4}\int_{\sig}\M\hat{A}_{a}
\end{equation}
 where
\begin{eqnarray*}
\hat{A}_{a}&=&4\hat{R_{0}}\Hb+2\der\D\Hb+2\kio\D\chi\Hb+2\chi\D\kio\Hb
\nonumber \\
	   &-&\D\kio\D\Hb-\D\chi\D\Hb-\kio\der\Hb-\chi\der\Hb
\end{eqnarray*}
or in a compact form using covariant derivatives
\begin{equation} \label{chiiie}
\hat{A}_{a}=4(\hat{R_{0}}-\Rko)\Hb+\D(\Delk+\Delko)\Hb
\end{equation}
 which thus becomes obviously globally defined and in a form that is quite
 similar to $\Gamma_3$.

 Here, the coefficients $\chi$	and $\chi_{0}$ are the superaffine connections
 related to the polydromic $\demi$-superdifferentials $\Psi$, $\Psi_{0}$
 according to (\ref{chi3b}) and (\ref{chi3c}).
 $\Rko$ is a particular SB-holomorphic superprojective connection related to
 $\chi$, $\chi_{0}$ by
 \begin{equation} \label{chiiif}
\Rko=\demi(\Delk\chi_{0}+\Delko\chi).
\end{equation}
Using the BRST transformation laws of fields given in section 1.7, we
obtain, after lengthy but straightforward calculations

\begin{equation} \label{chiiig}
s\Gamma_{a}[\Hb,\hat{R_{0}}]=k{\cal A}(\C,\Hb,\hat{R_{0}})+
k(K_{1}+K_{2}+K_{3})
\end{equation}
 where
\begin{eqnarray}\label{chiiig2}
 K_{1}&=& \fact\oint_{\dsig}\dl [(\hat{R_{0}}-\Rko)\C]	\nonumber\\
 K_{2}&=& -\facth\oint_{\dsig}\dlb (\Delk s\Hb+\Delko\Db\C+
 2\C\D\Db\chi+\D\C\Db\chi) \nonumber\\
 K_{3}&=& -\facth\oint_{\dsig}\dlb [\demi(\chi_{0}-\chi)\D(\D\Hb\D\C)
 -\chi_{0}\chi(\D\Hb\D\C)   \nonumber\\
      & &   -(\chi_{0}\D\chi+\der\chi-\chi_{0}\chi\D+\R)(\C\D\Hb+\Hb\D\C)]
 \end{eqnarray}

and ${\cal A}(\C,\Hb,\hat{R_{0}})$ is the globally defined integrated
anomaly given in (\ref{intre}).

First we wish to stress that the terms in eq.(\ref{chiiig2}) are separately
globally defined. This is obvious for $K_{1}$ since its integrand is the
product of the difference of two superprojective connections (i.e. a
superquadratic
differential) and $\C$ which transforms homogeneously. Concerning $K_{2}$ and
$K_{3}$ this property has to be checked by hand. While $K_{2}$ contains the
bosonic limit ( see ref.\cite{Zu1} eq.(2.15)), the fourth
term $K_{3}$ gathers the complications and novelties emerging from the
supersymmetric formulation. Note that all these terms are {\it a priori }
untractable
because they all involve multivalued fields. Nevertheless, we will show
later that these terms can be eliminated by adding other contributions to the
Polyakov action, thus leaving the anomaly only.

 Now let us continue our study of polydromic fields started in chapter I.

\subsubsection{Polydromic fields}

As it was shown in chapter I there are according to super Riemann-Roch
theorem $g$ holomorphic single-valued
$\demi-$superdifferentials which have each globally $(g-1)$ (counting
multiplicity) zeros on a compact SRS of genus $g$.
We shall consider in the following two kinds of these differentials;
we recall that the
holomorphic ones w.r.t. the SB-structure are denoted by $\eta$ and
the holomorphic objects in the $0-$structure by $\eta_{0}$.
In contrast, the multivalued \sdfs  $\Psi, \Psi_{0}$ are free of zeros, and
their general expression were given in (\ref{chi2s}) and (\ref{chi2s2}), see
section 1.4.2.\\
Now, to deal with these differentials we adopt the approach presented in
section 1.4.2, which consists in choosing a branch for
each of these differentials by working on a dissection of the SRS into its
polygon. As mentioned in the introduction to this chapter, we are
considering
a SRS with De Witt topology--recall that such SRS' are the most suitable for
a picture of a moving superstring. Now due to the fact this topology is
trivial in the $\theta$ direction, it is possible to borrow from the bosonic
case the method of cutting a Riemann surface into its fundamental domain.
Accordingly, we cut the SRS $\sig$  into
its polygon whose reduced domain is the polygon of the underlying Riemann
surface. This dissection has $4g$ pairwise opposite sides and will be
denoted by $\hat{\cal D}$, see section 1.4.2. This is assumed to contain
all zeros of
$\eta_{0}$ while excluding those of $\eta$. In the bosonic case, this was
shown to be always possible \cite{Zu1}, and even when zeros of the
(corresponding) bosonic differentials $\omega$ and $\omega_{0}$ overlap, the
total action is still continuous
\footnote{We are indebted to R. Zucchini for valuable discussions on this
subject and in particular on polydromic differentials\cite{Zu3}.}. Again
using the trivial topology of De Witt, one can repeat the same
demonstration on $\sig$. Next we cut infinitesimal
superloops ${\cal C}_{k}$ around the zeros $\hat P_{0k}$ of $\eta_{0}$ with
orientation opposite to that of the boundary $\partial\hat{\cal D}$
of $\hat{\cal D}$. These superloops are composed of their bodies
( ordinary circles in $\Sigma$ ) which define the corresponding orientation
and some ``Grassmann circle'' ${\cal C}_{\theta}$ ( defined below ) over
them. ${\cal C}_{\theta}$ represents the $\theta$ direction of
$\partial\hat{\cal D}$.\\

Next, in order to explain the line integrals in eqs.(\ref{chiiig2}) we
present our integration
procedure over a (compact) SRS with respect to which our solution to the
superconformal Ward identity (\ref{intre}) will be defined. Most importantly,
we
give the definition of the boundary of a superdomain $\hat{\cal D}$ in $\sig$,
and the analog of Stokes theorem to relate integration over superdomains on
$\sig$ and the one on their boundaries. \\

\subsubsection{Integration on $\sig$}

In our developments, the expression that we integrate over $\hat\Sigma$
is a $(\frac{1}{2},\frac{1}{2})-$superdifferential
which is, in general, meromorphic . More precisely, this expression
happens to be a function of singular objects like $\partial\log\eta_{0}$
( or $\D\log\eta_{0}$) inside a domain $\hat{\cal D}$ containing all zeros of
$\eta_{0}$. To perform explicitly the corresponding integral and in
particular the residue calculus, we first need an analog of Stokes theorem
and a consistent procedure of integration on the boundary
$\partial\hat{{\cal D}}$ of $\hat{{\cal D}}$. In fact, integration on
$\partial\hat{{\cal D}}$ reduces to the sum of integral over small
``circles'' ${\cal C}_{k}$ surrounding the zeros
$\hat{P}_{0k}=(z_{0k},\theta_{0k})$ of
$\eta_{0}$; the orientation of these circles being opposite to that of
$\partial\hat{{\cal D}}$.
Since the remaining integration path is a sequence of pairs of
geometrically coinciding but oppositely oriented arcs, this yields pairs of
mutually cancelling contributions when the integrand is single-valued.
Altogether, the only contribution comes from the
${\cal C}_{k}$'s. As explained above, these circles are composed of ordinary
circles in $\Sigma$ and some ``Grassmann circles'' ${\cal C}_{\theta}$
around the singular point
$\theta = \theta_{0},\; \bar\theta=\bar\theta_{0}$. ${\cal C}_{\theta}$
is defined in such a way that the identity
\begin{equation}\label{chiiii}
 \int_{{\cal C}_{\theta}}d\theta f(z,\theta,\bar z,\bar\theta)=
 (\partial_{\theta}f)(z,\theta_{0},\bar z,\bar\theta_{0})
\end{equation}
holds for every (locally) smooth function $f$ on $\hat\Sigma$.\\
Thus any integration over Grassmann numbers is performed over
${\cal C}_{\theta}$ instead of the whole space of Grassmann variables as is
done in the standard Berezin integration formalism. From (\ref{chiiii}) one
recovers the usual Berezin rules, which we now write on ${\cal C}_{\theta}$
\begin{equation}\label{chiiij}
\int_{{\cal C}_{\theta}}d\theta = 0,\hspace{1cm}
\int_{{\cal C}_{\theta}}d\theta(\theta-\theta_{0}) = 1,\hspace{1cm}
\int_{{\cal C}_{\theta}}d\theta(\bar\theta-\bar\theta_{0}) = 0.
\end{equation}

However, our integration procedure (\ref{chiiii}) marks a crucial departure
from
that of Berezin by the result
\begin{equation}\label{chiiik}
\int_{{\cal C}_{\theta}}d\theta(\theta-\theta_{0})
(\bar\theta-\bar\theta_{0}) = 0
\end{equation}
Indeed, in our point of view $\bar\theta$ is treated somehow as being linked
to $\theta$ on ${\cal C}_{\theta}$ and hence $\theta$ (or $\bar\theta$) can
not be taken out of the (line) integral over $\bar\theta$ (or $\theta$).
This super integration formalism is justified by the fact that it reproduces
the results obtained by the
component expansion while avoiding the well-known difficulties of this
procedure, especially when the superfields are singular. Obviously, the rules
(\ref{chiiii})-(\ref{chiiik}) are the same if ${\cal C}_{\theta}$ is a circle
around the origin $\theta_{0}=0,\bar\theta_{0}=0$. \\
In fact, this procedure basically amounts to ruling out the
$\bar\theta-$dependence from the expressions we integrate on a SRS, this is
not surprising since we are considering $(1,0)-$induced supergravity. On the
other hand, this is  equivalent to working in the gauge (\ref{chi1e}).\\
The rules above allow us to establish the super Stokes theorem
\footnote{See \cite{BL,VorZor} for other formulations of this theorem.}

\begin{equation}\label{chiiil}
\int_{\hat{\cal D}}\hat{d}\Phi = \oint_{\partial\hat{{\cal D}}}\Phi
\end{equation}
where the coboundary operator $\hat{d}$ is defined by its action on a
$(p/2,q/2)-$superdifferential $\Phi$ as follows
\footnote{$\hat{d}$ will be denoted $d_{+}$ or $d_{-}$ when $(p+q)$ is even
or odd respectively.}
\begin{equation}\label{chiiim}
\hat{d}\Phi=(\dl\D + (-1)^{(p+q)}\dlb\Db)\Phi.
\end{equation}
It is  straightforward to see that the operator $\hat{d}$ is nilpotent
as it must be, i.e.$\hat{d}^{2} = 0$.
For this, $\Phi$ is explicitly written as
$\Phi =\Phi_{\theta}(\dl)^{p}(\dlb)^{q}$
and the operators $\D$ and $\Db$ act directly on the coefficient
$\Phi_{\theta}$, thus leading to $d_{+}d_{-}=-d_{-}d_{+}=0$.\\
Now we wish to emphasize that this theorem is very crucial in that it allows
us to compute integrals in the superfield
formalism and thus spares us the generally cumbersome procedure of expanding
superfields in their components, especially when these are singular, since
there is in general no obvious way to tell which components exhibit
such or such singularities.

Now using these integration rules for Grassmann variables, the integral of a
meromorphic $(\frac{1}{2},0)-$ or $(0,\frac{1}{2})-$superdifferential $\Phi$
over the circles ${\cal C}_{k}$ is performed by using the
(generic) local behaviour around $\hat{P}_{0k}$ obtained from that of
$\eta_{0}$ in (\ref{chi2o}) \cite{HKMK1}
\begin{equation}\label{chiiin}
\Phi\sim \sum_{j}^{N}\frac{f_{j}}{(z-z_{0k}-\theta\theta_{0k})^{j}}
\end{equation}
where $N$ is the number of terms in $\Phi$, and
the coefficient functions $f_{j}$ are superholomorphic around
$\hat{P}_{0k}$, i.e. they do not depend on $(\bar z,\bar\theta)$ inside an
open neighbourhood of $\hat{P}_{0k}$ contained in ${\cal C}_{k}$. Then we get
the final result with the help of the super Cauchy theorems\cite{Fried}
\begin{eqnarray}\label{chiiio}
\fact\oint_{{\cal C}_{k}}\dl f(z,\theta)(\theta-\theta_{k})
(z-z_{k}-\theta\theta_{k})^{-n-1}
 &=& \frac{1}{n!}\partial_{z_{k}}^{n}D_{\theta_{k}}f(z_{k},\theta_{k})
 \nonumber\\
\fact\oint_{{\cal C}_{k}}\dl f(z,\theta)(z-z_{k}-\theta
\theta_{k})^{-n-1} &=& \frac{1}{n!}\partial_{z_{k}}^{n} f(z_{k},\theta_{k}),
\end{eqnarray}

\vspace{1cm}
After these two digressions devoted to developing two of the building blocks
in our approach, we now resume the construction of a Polyakov action
started with $\Gamma_{a}$ in (\ref{chiiid}). It is clear that additional
terms are required in order to cancel the terms $K_{1},K_{2},K_{3}$ in
(\ref{chiiig}), and thus
to get the anomaly only. The next obvious term, say $\Gamma_b$, is simply the
supersymmetric copy of $\Gamma_{4}[\mu]$ in (\ref{chiiib}), that is

\begin{equation}\label{chiiip}
\Gamma_{b}[\Hb]=\fact\oint_{\dsig}\ln(\Psi/\eta_{0})\hat{d}
\ln(\eta/\Psi_{0})\equiv\fact\oint_{\dsig}\hat{A}_{b}
\end{equation}
Obviously the integrand here is globally defined as it involves only ratios
of superdifferentials, i.e. invariant functions.\\
The action of the BRST operator on $\Gamma_b$ results in many terms, see
details in {\bf IV}. One of them is
\begin{eqnarray*}
I=\fact\oint_{\dsig}(\R-R_{\zeta_{0}})\C
\end{eqnarray*}
which is easily computed by following the integration procedure described
in the previous section. This yields
\begin{equation}\label{chiiiq}
I=-\frac{1}{4}\sum_k\alpha_{0k}\Delta_{\zeta}\C(\hat{P}_{0k})
=-s\left[\demi\sum_k\alpha_{0k}\ln(\eta/\Psi_{0})(\hat{P}_{0k})\right].
\end{equation}
Here we use the fact that the $s$-variation commutes with the evaluation at a
point $\hat P\in\hat\Sigma$.\\
This suggests that the term on which $s$ acts in
(\ref{chiiiq}), call it $\Gamma_c$, must be added to $\Gamma_a+\Gamma_b$ in
order to cancel $I$. $\Gamma_c$ is on the other hand the obvious
generalization of $\Gamma_5$ in (\ref{chiiib}), but here it comes out from
explicit calculations anyway.\\
Two other terms in $s\Gamma_b$ cancel with $K_1$ and $K_2$ in (\ref{chiiig}).
However, this is not finished because some of the remaining terms in
$s\Gamma_b$ contain polydromic fields and hence cannot be integrated,
but they instead require extra terms to cancel them out. In addition,
other terms
in $s\Gamma_b$ and $K_3$ in (\ref{chiiig}) are more difficult to deal with
as they are quadratic in $\zeta$'s and $\chi$'s. This makes it clear that we
still need an additonal contribution to $\Gamma_a+\Gamma_b+\Gamma_c$ in order
to get rid of all these terms. More precisely we add the following term

\begin{equation} \label{chiiir}
\Gamma_{3}[\Hb]=\frac{1}{4i\pi}\oint_{\dsig}\dlb [(\zeta\zeta_{0}
		-\chi\chi_{0}-2\zeta\chi)\Hb-\demi(\chi+\chi_{0}-\zeta
		 -\zeta_{0})\D\Hb]
\end{equation}

It is an easy matter to check that the integrand of this expression is
globally defined, and that when expanded in components \cite{AK1} it
disappears in the bosonic limit. In {\bf IV} we show how this contribution
indeed rids us of all the above-mentioned undesirable terms.

To recapitulate, we have shown that a general Polyakov action on a SRS of
genus $g$ contains three kinds of terms, an integral on $\sig$, a line
integral and a residue contribution
\begin{eqnarray} \label{chiiis}
\Gamma_{a}[\Hb,\hat R_{0}]&=&\frac{1}{4\pi}\int_{\hat\Sigma} \M
[4(\hat R_{0}-\Rko)\Hb+\D(\Delk+\Delko)\Hb]  \nonumber \\
\Gamma_{b}[\Hb]&=&\fact\oint_{\dsig}\ln(\Psi/\eta_{0})\hat{d}
\ln(\eta/\Psi_{0})\nonumber \\
\Gamma_{c}[\Hb]&=&\frac{1}{4i\pi}\oint_{\dsig}\dlb
\{ (\zeta\zeta_{0}-\chi\chi_{0}-2\zeta\chi)\Hb-
\demi(\chi+\chi_{0}-\zeta-\zeta_{0})\D\Hb\} \nonumber\\
\Gamma_{d}[\Hb]&=&\demi\sum_k\alpha_{0k}\ln(\eta/\Psi_{0})(\hat P_{0k}).
\end{eqnarray}

This solves the superconformal Ward identity (\ref{intre}) on a SRS of higher
genus and it is holomorphic w.r.t. the super Beltrami differential $H$.
The genus which characterizes the SRS appears explicitly in
$\Gamma_{d}$, since $\sum_{k=1}^{N}\frac{\alpha_{0k}}{2}=g-1$, where $N$ is
the number (without counting multiplicity) of zeros of $\eta_{0}$. See below
for other properties of this action.

\subsection{Non-uniqueness of the Polyakov action on a SRS}

Here, we wish to emphasize that by construction this solution is, as
discussed in the introduction, not unique
due to the presence of zero modes, i.e. it is only defined up to addition
of an arbitrary functional $\hat\Phi(H)$ satisfying the
conditions~~$\delta \hat\Phi(H)/\delta \bar H=0,~ s\hat\Phi(H)=0$.
In fact we can
even go further as in the bosonic case and start with the action $\Gamma_{a}$
in (\ref{chiiis}) and replace the multivalued fields
$\Psi$ and $\Psi_{0}$ by the single-valued ones $\eta$ and $\eta_{0}$
respectively, this results in an expression which can be shown to be the
first term for a second solution to the superconformal Ward identity
(\ref{intre}).
More precisely, let us denote the integrand in the resulting expression by
$\hat A_{\alpha}$ and then by using the holomorphy equations satisfied by
$\Psi, \Psi_{0}, \eta, \eta_{0}$
\begin{equation} \label{chiiit}
\Db\ln\Psi=\demi\Delk\Hb  \hspace{.5cm} , \hspace{.5cm} \Db\Psi_{0}=0
\hspace{.5cm} , \hspace{.5cm} \Db\ln\eta=\demi\Del\Hb  \hspace{.5cm}
, \hspace{.5cm}  \Db\eta_{0}=0
\end{equation}

we get
\begin{eqnarray}\label{chiiiu}
\hat{A}_{a}-\hat{A}_{\alpha}&=&-4\{\Db[\ln(\Psi/\eta_{0})
\D\ln(\eta/\Psi_{0})]
+ \D[\ln(\Psi/\eta_{0})\Db\ln(\eta/\Psi_{0})]\} \\ \nonumber
&-&\D\{ (\zeta\zeta_{0}-\chi\chi_{0}-2\zeta\chi)\Hb-
\demi(\chi+\chi_{0}-\zeta-\zeta_{0})\D\Hb\}
\end{eqnarray}
where $\hat{A}_{a}$ is given in (\ref{chiiie}).
Next we integrate this expression over $\hat\Sigma$ which is now seen as the
disjoint union of the domains $\hat{{\cal D}}(\eta_{0})$ and
$\hat{{\cal D}}(\eta)$ that contain all zeros of $\eta_{0}$ and $\eta$
respectively, and then use the Stokes theorem (\ref{chiiil}) to obtain the
identity
\begin{eqnarray}\label{chiiiv}
\Gamma_{\alpha} &=& \Gamma_{a} + \fact\oint_{\dsig}\ln(\Psi/\eta_{0})\hat{d}
\ln(\eta/\Psi_{0}) + \fact\oint_{\partial\hat{{\cal D}}(\eta)}
\ln(\Psi/\eta_{0})\hat{d}\ln(\eta/\Psi_{0})\nonumber \\
&+&\frac{1}{4i\pi}\oint_{\partial\hat\Sigma}\dlb
\{ (\zeta\zeta_{0}-\chi\chi_{0}-2\zeta\chi)\Hb-
\demi(\chi+\chi_{0}-\zeta-\zeta_{0})\D\Hb\}
\end{eqnarray}

Note that in the third term we have integrable singularities and thus
by following our integration procedure we obtain
$$\fact\oint_{\partial\hat{{\cal D}}(\eta)}
\ln(\Psi/\eta_{0})\hat{d}\ln(\eta/\Psi_{0})=
-\demi\sum_j\alpha_{j}\ln(\Psi/\eta_{0})(\hat P_{j})$$

Consequently, the sum $\Gamma_{a}+\ldots+\Gamma_{d}$ is equal to the sum of
the following terms
\begin{eqnarray}\label{chiiiw}
\Gamma_{\alpha}[\Hb,\hat R_0]&=&\frac{1}{4\pi}\int_{\hat\Sigma} \M
[4(\hat R_{0}-R_{\zeta_{0}})\Hb+\D(\Delta_{\zeta}+\Delta_{\zeta_{0}})\Hb]
\nonumber \\
\Gamma_{\beta}[\Hb]&=&\demi\sum_k\alpha_{0k}\ln(\eta/\Psi_{0})(\hat P_{0k})
+\demi\sum_j\alpha_{j}\ln(\Psi/\eta_{0})(\hat P_{j}) \nonumber \\
\Gamma_{\gamma}[\Hb]&=&\frac{-1}{4i\pi}\oint_{\hat{{\cal D}}(\eta)}\dlb
\{ (\zeta\zeta_{0}-\chi\chi_{0}-2\zeta\chi)\Hb-
\demi(\chi+\chi_{0}-\zeta-\zeta_{0})\D\Hb\}
\end{eqnarray}
As mentioned above, this solution can be seen as the supersymmetric
generalization of the solution in (\ref{chiiia}). Now let us show that by
means of this solution we can make contact with the supertorus.\\

\subsection{Relating the Polyakov action on a SRS and on the supertorus}

One of the advanges of the solution (\ref{chiiiw}) over that in
(\ref{chiiis}) is the
fact that it can be easily related to the Polyakov action (\ref{chiil}) we
constructed on the supertorus. Indeed, the same procedure as in section 3.3
yields here the following identity
\begin{eqnarray}\label{chiiix}
\hat A_{a} &=& A_{ST} - D[\D\Hb\D\ln(\eta/\eta_{0}) -
2\Hb\D\ln\eta\D\ln\eta_{0}] + 4\D\Db\ln\eta \nonumber \\
&\equiv& A_{ST} + I_{1} + I_{2}
\end{eqnarray}
which holds on a SRS. Then using the fact that $\eta_{0}$ is holomorphic in
the reference structure i.e., $\Db\eta_{0}=0$, we can rewrite $I_{2}$ as

$$I_{2}= 4 \D\Db\ln(\eta/\eta_{0})$$
thus yielding a well-defined expression since $\eta/\eta_{0}$ is now a
function. $I_2$ is therefore a total derivative of a single-valued,
well-defined and singularity free  $\demi-$superdifferential
$\D\ln(\eta/\eta_{0})$, and hence vanishes upon integrating on small
circles on the supertorus.\\
Similarly, $I_{1}$ is a total derivative of a single-valued and
non-singular  $(0,\frac{1}{2})-$superdifferential, since the expression in
brackets in
(\ref{chiiix}) transforms with the factor $\exp{(-\bar\varpi)}$ under
conformal change of coordinates. Thus the integral of $I_1$ over the
supertorus vanishes.
Therefore the restriction of $\hat A_a$ onto the supertorus gives exactly
$A_{ST}$. \\
Now as the differentials on the supertorus have according to the super
Riemann-Roch theorem no zeros, $\Gamma_{\beta}$ vanishes trivially since
$\alpha_{0k}=\alpha_{j}=0$. As to $\Gamma_{\gamma}$ the reasoning is as
follows. Due to the fact that there is a unique holomorphic superdifferential
(in a given structure) on the supertorus, $\Psi$ and $\Psi_{0}$ become
proportional to $\eta$ and
$\eta_{0}$ respectively, and the corresponding factors are multivalued
functions. However, these must be holomorphic on the whole torus
so that this restriction holds everywhere thereon, which
implies that they are constants. In this
case $\chi$ and $\chi_{0}$ reduce exactly to $\zeta$ and $\zeta_{0}$, and
thereby $\Gamma_{\gamma}$ vanishes identically on the supertorus.\\
Here again we see that the supertorus can be related through the
Polyakov action to the general case of a SRS of higher genus.

\subsection{Stress-energy tensor and OPE on a SRS}

As discussed in the introduction, the existence of singular and polydromic
fields on a SRS, together with the lack of a well-established superfield
formalism, makes it
rather difficult to compute the stress-energy tensor and OPE from the actions
(\ref{chiiis}) and (\ref{chiiiw}). Nevertheless, we know that from the
geometrical point of view
the stress-energy tensor is a (super) projective connection times the
central charge of the (super) Virasoro algebra it generates. In addition,
as was often stressed above, the physics
underlying the Polyakov action on a (super) Riemann surface depends on the
conformal geometry represented by the (super) Beltrami differential and a
background (super) projective
connection. Thus, since the term containing the latter is linear in the
Beltrami differential, we expect this connection to appear only in the stress-
energy tensor. On the supertorus this yields the
exact result (\ref{chiix}), and on the supercomplex plane we have,
{}~${\cal T}(a)=-2k \hat R_{\zeta}(a)$.

However, on a (super) Riemann surface this is not the whole story since
this result would be singular thereon as is the particular connection
$\hat R_{\zeta}$ (see e.g.(\ref{chiix}) or Ref.\cite{Zu1}). So in order to
get a regular stress-energy tensor we expect an additional term which would
compensate for the singularities of $\hat R_{\zeta}$. In addition, still
another contribution comes from the $\hat \Phi-$indetermination  in the action.
\\
Thus taking all this into account, a general expression for the stress-energy
tensor of the Polyakov action (\ref{chiiiw}) on a compact SRS can be written
in a form similar to the bosonic case \cite{Zu1}, that is
\begin{equation}\label{chiiiy}
{\cal T}(z,\theta)=\frac{\delta\Gamma_{SRS}}{\delta H(z,\theta)}
=k\left\{(\hat R_{0}-\hat R_{\zeta})+\frac{\delta\hat\Phi(H)}{\delta H}+
S(H)\right\}.
\end{equation}
Here $\hat R_{\zeta}=-\partial\D\ln\eta-\partial\ln\eta\D\ln\eta$~ has poles
where $\eta$ vanishes. To be more precise about the term $S(H)$, let
$\Phi^{\demi,0}$ denote as in section 1.5 either $\eta$ or $\Psi$. Then
$S(H)$ is singular since it is a function of $\delta\Phi^{\demi,0}/\delta H$
which is singular due to the fact that $\Phi^{\demi,0}$ is
holomorphic in the SB-structure on a SRS.\\
For $\hat{P}, \hat{P}^{'}\in\hat\Sigma$,
{}~$\delta\Phi^{\demi,0}(\hat P)/\delta H(\hat{P}^{'})$
behaves like a $\demi-$superdifferential at
$\hat{P}\neq\hat{P}^{'}$, i.e. it satisfies the holomorphy equation
\begin{equation}\label{chiiiz}
\left[
\bar{D}_{\hat P}-H(\hat P)\partial_{\hat P}-\demi(\partial H)(\hat P)
+\demi(DH)(\hat P)D_{\hat P}
\right]\frac{\delta\Phi^{\demi,0}(\hat P)}{\delta H(\hat{P}^{'})}=0,
\end{equation}
while at $\hat{P}^{'}\neq\hat{P}$, it behaves like a quadratic (or
$\frac{3}{2}-$) superdifferential satisfying
\begin{equation}\label{chiiiz1}
\left[
\bar{D}_{\hat{P}^{'}}-H(\hat{P}^{'})\partial_{\hat{P}^{'}}
-\frac{3}{2}(\partial H)(\hat{P}^{'})
+\demi(DH)(\hat{P}^{'})D_{\hat{P}^{'}}
\right]\frac{\delta\Phi^{\demi,0}(\hat{P})}{\delta H(\hat{P}^{'})}=0.
\end{equation}
Now if we define $\delta\Phi^{\demi,0}(\hat P)/\delta H(\hat{P}^{'})$ as being
the action of the functional operator~~ $t(\hat{P}^{'})\equiv\frac{\delta}
{\delta H(\hat{P}^{'})}$, that we call the``stress-energy operator''
\footnote{As distinguished from the stress-energy tensor ${\cal T}(\hat P)
\equiv t(\hat P)\Gamma[H]$ coupled to the super Beltrami differential $H$ in
the action $\Gamma[H]$.} on a superconformal field $\Phi^{\demi,0}(\hat P)$
of conformal weights $(\demi,0)$, we can write it as the
OPE of $t(\hat{P}^{'})$ with the field $\Phi^{\demi,0}(\hat P)$.
To write down the corresponding expression let us use the superprojective
structure $\{(\hat Z,\hat\Theta)\}$. Correspondingly, we have
$$T(\hat{P}^{'})\equiv T(\hat Z(\hat{P}^{'}),\hat\Theta(\hat{P}^{'}))=
\Lambda^{-3/2}t(\hat P)$$
$\Lambda$ was defined in (\ref{chi2g}), and here it becomes
{}~$\Lambda=(\D\hat\Theta)^{2}$ in the gauge $H^{z}_{\theta}=0$; and
$$\tilde{\Phi}^{\demi,0}(\hat{P}^{'})\equiv
\tilde{\Phi}^{\demi,0}(\hat Z(\hat{P}^{'}),\hat\Theta(\hat{P}^{'}))
=\Lambda^{-\demi}\Phi^{\demi,0}(\hat P)$$
Finally, following similar calculus as in the bosonic case \cite{Zu1}, we can
write
\begin{eqnarray}\label{chiiiz2}
\frac{\delta\Phi^{\demi,0}(\hat{P})}{\delta H(\hat{P}^{'})}&=&
T(\hat{P}^{'})\tilde{\Phi}^{\demi,0}(\hat{P})\sim
\Lambda^{\demi}(\hat{P})\Lambda^{3/2}(\hat{P}^{'})
\left\{
\frac{\demi~\tilde{\Phi}^{\demi,0}(\hat P)
(\hat\Theta(\hat{P}^{'})-\hat\Theta(\hat{P}))}
{(\hat{Z}(\hat{P}^{'})-\hat{Z}(\hat P)-
\hat{\Theta}(\hat{P}^{'})\hat{\Theta}(\hat{P}))^{2}}\right.\nonumber\\
&&\nonumber\\
&+&\left.\frac{\demi\;D_{\hat\Theta}\tilde{\Phi}^{\demi,0}(\hat P)}
{(\hat{Z}(\hat{P}^{'})-\hat{Z}(\hat P)-\hat{\Theta}(\hat{P}^{'})\hat{\Theta}
(\hat{P}))}+\frac{(\partial_{\hat Z}\tilde{\Phi}^{\demi,0})(\hat P)
(\hat{\Theta}(\hat{P}^{'})-\hat{\Theta}(\hat{P}))}
{(\hat{Z}(\hat{P}^{'})-\hat{Z}(\hat P)-
\hat{\Theta}(\hat{P}^{'})\hat{\Theta}(\hat{P}))}
\right\}\nonumber\\
&&\nonumber\\
&+&R.T.
\end{eqnarray}
This formula expresses the main properties of
$\delta\Phi^{\demi,0}(\hat P)/\delta H(\hat{P}^{'})$ reflected by the
holomorphy equations given above. The poles appearing in it are expected to
compensate those of $\hat R_{\zeta}$ so that ${\cal T}(z,\theta)$ in
(\ref{chiiiy}) becomes regular and thus single-valued. Moreover,
${\cal T}(z,\theta)$ must satisfy the following integrability condition
\begin{equation}\label{chiiiz3}
t(\hat P){\cal T}(\hat{P}^{'})=t(\hat{P}^{'}){\cal T}(\hat{P})
\end{equation}
which could be used in addition to modular invariance to study the physical
properties of the Polyakov action (\ref{chiiiw}).\\
The computation of $N-$point functions should in principle follow the same
formalism, but a rigorous study is still needed.

\subsection{Conclusion}
Despite all these difficulties, we have been able to construct the effective
action for the conformal $N=1$ $2D-$induced supergravity on a super Riemann
surface of genus $g>1$, which is holomorphic w.r.t. the super Beltrami
differential and integrates the superdiffeomorphism anomaly thereon. This has
been given in two forms starting with two sets of fields with different
monodromy
properties. The solution that starts with single-valued fields has been
related to the action found on the supertorus. Using this solution we gave the
general expression for the stress-energy tensor and the OPE of the
stress-energy operator with $\demi-$superdifferentials. However, we still
need to perform the same calculations with the action that starts with
polydromic fields.

In future, we would like to work out the computation of $N-$point
functions. We believe that this would be possible by generalizing the method
used on the supertorus, however this requires rigorous study of the properties
of the
superprime form discussed in section 1.3.\\
There are many other issues that deserve serious work, namely the modular
invariance of these solutions and their pertinence to resumming the
perturbative series provided by the renormalized field theory as an iterative
solution to the superconformal Ward identity (\ref{intre}). This would
provide a generalization of the similar work done on the supertorus in
{\bf III}.

\section{Concluding comments}
In the present work we have treated the last and most difficult cases in the
formulation of the chiral-gauge 2-dimensional gravity in the Polyakov
formalism on Riemann surfaces.\\
Indeed we have constructed regular and single-valued Polyakov actions on the
supertorus and a super Riemann surface of higher genus. These integrate the
superdiffeomorphism anomaly thereon and are holomorphic with respect to the
super Beltrami differential that represents the graviton-gravitino gauge
doublet. To perform this task on a super Riemann surface we were led to
develop some new algebraic and geometrical appropriate tools such as the
notion of polydromy and branch-cutting. We have also defined a new integration
method for Grassmann variables including our formulation of Stokes
theorem using a new boundary operator; the $\bar\partial-$
Cauchy kernel has been constructed on the supertorus and only discussed on
higher genus super Riemann surfaces. \\
Then by computing the energy-momentum tensor and operator product expansions
to the third order, we have shown that the Polyakov action on the supertorus
( and the supercomplex plane ) resums, at least to this order, the
perturbative series provided by the
renormalized field theory. On a super Riemann surface, we have given the
general expression for the energy-momentum tensor and its operator product
expansions with $\demi-$superdifferentials which are the building blocks of
our work. However, we have not succeeded yet in computing higher order operator
product expansions in this case.

Among the open problems that we would like to investigate in future,
we can mention the following :

1) The need to compute $N-$point functions and thereby prove to all orders
that the Polyakov action found here resums the perturbative series, at least
on the supertorus.

2) Perform the same calculus on a super Riemann surface. This requires a more
elaborate study of the super prime form, e.g. the one we have suggested.

3) Study the modular invariance and exploit the integrability condition
(\ref{chiiiz3}) in order to fix the ambiguity in the Polyakov action on a
( super ) Riemann surface. This will require, among other things, a
formulation of our setting in the other spin structures.

\end{document}